\documentclass{agujournal2019}
\usepackage{url} 
\usepackage{lineno}
\usepackage[inline]{trackchanges} 
\usepackage{soul}
\usepackage{siunitx}
\usepackage{amssymb}
\usepackage{amsmath}
\usepackage{float}
\usepackage{comment}
\journalname{JGR: Space Physics}

\begin{document}

%
%


\title{Reconstruction of polarization properties of whistler waves from two magnetic and two electric field components: application to Parker Solar Probe measurements}

%
%




\authors{L. Colomban\affil{1}, O. V. Agapitov\affil{2,7}, V. Krasnoselskikh\affil{1,2}, M. Kretzschmar\affil{1}, T. Dudok de Wit\affil{1,8}, S. Karbashewski\affil{2}, F. S. Mozer\affil{2}, J. W.  Bonnell\affil{2}, S. Bale\affil{2,3}, D. Malaspina\affil{4,5},  N. E. Raouafi\affil{6}}

\affiliation{1}{LPC2E, CNRS/University of Orléans/CNES, 3A avenue de la Recherche Scientifique, Orléans, France}
\affiliation{2}{Space Sciences Laboratory, University of California, Berkeley, CA 94720-7450, USA}
\affiliation{3}{Physics Department, University of California, Berkeley, CA, United States}
\affiliation{4}{Laboratory for Atmospheric and Space Physics, University of Colorado, Boulder, CO 80303, USA}
\affiliation{5}{Astrophysical and Planetary Sciences Department, University of Colorado, Boulder, CO 80303, USA}
\affiliation{6}{Applied Physics Laboratory, Johns Hopkins University, Laurel, MD, United States}
\affiliation{7}{Astronomy and Space Physics Department, National Taras Shevchenko University of Kyiv, Kyiv, Ukraine}
\affiliation{8}{International Space Science Institute, ISSI, Bern, Switzerland}




\correspondingauthor{L.Colomban}{lucas.colomban@cnrs-orleans.fr}

\def\cvk#1{\textcolor{blue}{(\bf Volodya: \bf#1)}}
\def\cmk#1{\textcolor{magenta}{(\bf matthieu: \bf#1)}}
\def\clk#1{\textcolor{orange}{(\bf Lucas: \bf#1)}}
\def\clo#1{\textcolor{green}{(\bf Oleksiy: \bf#1)}}




\begin{keypoints}

\item We present a method to determine whistler wave polarization without a magnetic field component (Parker Solar Probe regime after March 2019).

\item This allows us to expand whistler wave statistical databases; this is an essential step to better understanding wave-particle interactions. 

\item We demonstrate that this method applies to 80\% of whistler waves observed in burst-mode data from the first encounter.


\end{keypoints}

%
%

%
%


\begin{abstract}

The search-coil magnetometer (SCM) aboard Parker Solar Probe (PSP) measures the 3 Hz to 1 MHz magnetic field fluctuations. During Encounter 1, the SCM operated as expected; however, in March 2019, technical issues limited subsequent encounters to two components for frequencies below 1 kHz. Detrimentally, most whistler waves are observed in the affected frequency band where established techniques cannot extract the wave polarization properties under these conditions. Fortunately, the Electric Field Instrument (EFI) aboard PSP measures two electric field components and covers the affected bandwidth. We propose a technique using the available electromagnetic fields to reconstruct the missing components by neglecting the electric field parallel to the background magnetic field. This technique is applicable with the assumptions of (i) low-frequency whistlers in the plasma frame relative to the electron cyclotron frequency; (ii) a small propagation angle with respect to the background magnetic field; and (iii) a large wave phase speed relative to the cross-field solar wind velocity. Critically, the method cannot be applied if the background magnetic field is aligned with the affected SCM coil. We have validated our method using burst mode measurements made before March 2019. The reconstruction conditions are satisfied for 80\% of the burst mode whistlers detected during Encounter 1. We apply the method to determine the polarization of a whistler event observed after March 2019 during Encounter 2. Our novel method is an encouraging step toward analyzing whistler properties in affected encounters and improving our understanding of wave-particle interactions in the young solar wind.

\end{abstract}


%
%

\section{Introduction}
Parker Solar Probe (PSP) was launched in August 2018 to collect measurements of plasma parameters and electromagnetic fields in the inner heliosphere below 50 solar radii ($R_{\odot}$) offering unique opportunities to study \textit{in situ} the young solar wind \cite{fox_solar_2016, bale_highly_2019,howard_near-sun_2019,kasper_alfvenic_2019,mccomas_probing_2019,raouafi_parker_2023}. The first solar encounter had its perihelion at 35.7 $R_{\odot}$; subsequent perihelia over the seven-year mission will drop closer to the Sun, eventually reaching a heliocentric distance of about 10 $R_{\odot}$ in mid-2025. The mission addresses two fundamental problems in space physics: coronal plasma heating and the acceleration of solar wind plasmas. In both problems, wave-particle interactions involving magnetohydrodynamic (MHD) and kinetic-scale waves are known to play an important role; one of the widely studied waves in the latter category is the whistler wave. \\

Whistlers are right-hand polarized electromagnetic modes observed between the lower hybrid frequency ($f_{\rm lh}$) and electron cyclotron frequency $f_{\rm ce}$ \cite{gurnett_plasma_1977,neubauer_fast_1977,lacombe_whistler_2014,chust_observations_2021,kretzschmar_whistler_2021}. Common whistler generation mechanisms are heat flux instability \cite{gary_electron_1975,tong_whistler_2019,lopez_particle--cell_2019}, fan instability \cite{kadomtsev_electric_1968,boskova_fan_1992,krafft_interaction_2003,vasko_whistler_2019,verscharen_self-induced_2019}, temperature anisotropy instability \cite{sagdeev_instability_1960,lazar_instability_2011,lazar_electromagnetic_2013,lazar_electromagnetic_2018,jagarlamudi_whistler_2020,vasko_quasi-parallel_2020}, and electron populations trapped in magnetic field inhomogeneities \cite{agapitov_sunward-propagating_2020,karbashewski_whistler_2023} often associated with boundaries of localized magnetic field deflections that are called switchbacks \cite{bale_highly_2019,kasper_alfvenic_2019,dudok_de_wit_switchbacks_2020,krasnoselskikh_localized_2020, agapitov_flux_2022}. These waves have been studied by several missions, such as Helios \cite{jagarlamudi_whistler_2020}, Cluster \cite{lacombe_whistler_2014}, Artemis \cite{stansby_experimental_2016, tong_statistical_2019}, Solar Orbiter \cite{kretzschmar_whistler_2021,chust_observations_2021}, STEREO \cite{breneman_observations_2010,cattell_narrowband_2020} and PSP \cite{ agapitov_sunward-propagating_2020, cattell_narrowband_2020, cattell_parker_2021, cattell_parker_2022, jagarlamudi_whistler_2021, dudok_de_wit_first_2022,froment_whistler_2023,karbashewski_whistler_2023}. As more observations are made about the dynamics of the solar wind, whistler waves have emerged as a strong candidate for interacting efficiently with solar wind electron populations \cite{gary_electron_1975,gary_solar_1977,scime_regulation_1994, gary_electron_1999, vocks_generation_2003,kajdic_suprathermal_2016}. \\


Electrons in the solar wind are accurately described by three distinct populations: a bulk thermal component with close to Maxwellian distributions and two suprathermal fractions called the Strahl and halo. The halo population is quasi-isotropic \cite{feldman_solar_1975,feldman_characteristic_1978,lazar_characteristics_2020} and is often represented by Kappa distributions \cite{scudder_causes_1992,scudder_why_1992,maksimovic_kinetic_1997,lazar_towards_2015}. The Strahl is a beam of high-energy electrons that follows the magnetic field lines, propagating away from the Sun \cite{rosenbauer_preliminary_1976,rosenbauer_survey_1977,pilipp_characteristics_1987}. 
The relative proportions of the two suprathermal populations are observed to evolve with radial distance from the Sun. Notably, the fractional density of halo electrons increases with distance while the Strahl distribution broadens \cite{hammond_variation_1996,graham_evolution_2017,bercic_scattering_2019} and its fractional density decreases with distance \cite{maksimovic_radial_2005, stverak_radial_2009,graham_evolution_2017,halekas_electrons_2020, halekas_electron_2021}. These observations suggest there are mechanisms, such as wave-particle interactions involving whistler waves, responsible for scattering the beam-like Strahl electrons into a more isotropic halo distribution. Several polarization properties determine the efficiency of whistler wave interactions with the Strahl and halo suprathermal electron populations. For example, two important parameters are the direction of propagation with respect to the Sun (i.e., sunward or anti-sunward propagation), and the wave normal angle (WNA), $\theta$, which is the angle between the wave vector, $\mathbf{k}$, and the background magnetic field, $\mathbf{B_\mathrm{0}}$. Sunward propagating whistlers scatter the Strahl population in the pitch-angle space more efficiently (up to two orders of magnitude greater) than the anti-sunward propagating waves \cite{vocks_electron_2005, saito_all_2007, sarfraz_combined_2020, cattell_modeling_2021}. The situation becomes more complicated if anti-sunward whistler waves have high WNA and thus provide efficient scattering \cite{artemyev_storm-induced_2013, artemyev_oblique_2016,vasko_whistler_2019, roberg-clark_scattering_2019, micera_particle--cell_2020, micera_role_2021, cattell_modeling_2021, cattell_parker_2022}. This makes the polarization properties of whistler waves observed by PSP one of the key factors for the quantification of the wave-particle interaction effects in the solar wind.

To date, a wide array of whistler observations made by PSP have been reported on. In previous studies, the polarization properties of whistler waves in the young solar wind were inferred from the three SCM components during Encounter 1 \cite{agapitov_sunward-propagating_2020, cattell_narrowband_2020, cattell_parker_2021, cattell_parker_2022, dudok_de_wit_first_2022, froment_whistler_2023,karbashewski_whistler_2023}.
The statistical study of whistler properties by \citeA{froment_whistler_2023} revealed that most of the whistler wave packets recorded during Encounter 1 were quasi-parallel to the background magnetic field: 97$\%$ had  WNA between 0 and 25$^\circ$. Whistler waves were observed in the frequency range from the local lower hybrid frequency $f_{\rm lh}$ up to $0.2 f_{\rm ce}$. The observed oblique whistlers (with WNA $> 45 ^\circ$) tend to have lower frequencies than the quasi-parallel whistlers. The sunward propagating whistler waves, both quasi-parallel and oblique waves, were often collocated with short-lived magnetic dips observed at switchback boundaries; this indicates a possible generation of whistlers in these structures \cite{agapitov_sunward-propagating_2020, froment_whistler_2023,karbashewski_whistler_2023}. These waves tend to be detected at frequencies that are lower than those for waves that are not collocated with magnetic dips \cite{froment_whistler_2023}. Another statistical study by \citeA{cattell_parker_2022}, on the basis of electric field measurements from the first nine encounters, showed that below the heliocentric distance of 100 $R_{\odot}$ whistler wave frequencies in the spacecraft frame were below $0.2f_{\rm ce}$ with the tendency to decrease below $0.1f_{\rm ce}$ when approaching the Sun closer than 50 $R_{\odot}$. 
To further elucidate the impact of whistler waves on the suprathermal electrons it is necessary to extend the statistics reported by \citeA{froment_whistler_2023} for Encounter 1 (the only one available with a full set of SCM magnetic measurements \cite{dudok_de_wit_first_2022}) to the later encounters and update the statistics of whistler waves presented by \citeA{cattell_parker_2021} and \citeA{cattell_parker_2022} with the wave polarization parameters.

A change in the response in one of the SCM components: $B\rm w_{\rm u}$ of the SCM reference frame,  $(\textbf{u},\textbf{v},\textbf{w})$ \cite{malaspina_digital_2016}, appeared after March 2019. Here and in the following $\vec{B}\rm w$ and $\vec{E}\rm w$ represent the wave magnetic and electric perturbations, respectively. This anomaly considerably reduces the amplitude of the $B\rm w_{\rm u}$ component in the frequency range of whistler waves (typically, below 400 \si{\hertz}) and also affects its phase. This makes it impossible to unambiguously determine the polarization properties using the three components of the magnetic field and the two components of the electric field. The inability to determine the whistler wave properties beyond Encounter 1 has motivated us to propose a novel technique for reconstructing the whistler wave polarization parameters. The technique uses the two components of the SCM that are available together with the two electric field components recorded by the EFI. 
It can be noted that the STEREO spacecraft have electrical antennas that measure 3 components of the electric field but do not have SCMs. \citeA{breneman_observations_2010} therefore developed a method based on the whistler wave cold dispersion relation and the ratio of transverse electric field components to determine whistler WNAs. This method is not designed to reconstruct electromagnetic field components and is different from the one proposed here. In addition, in the case of \citeA{breneman_observations_2010} the absence of magnetic field fluctutaions measurements makes it impossible to determine the direction of propagation. 

In the following, we present the data used (Section \ref{data}), our reconstruction method (Section \ref{The reconstruction technique}) and its range of applicability (Section \ref{Approximations}). We then detail the reconstruction of three whistler wave packets from Encounter 1 (Sections \ref{Case1}, \ref{Case 2}, \ref{case3}). We finally discuss the applicability of the method to other encounters (Section \ref{Discussion on the applicability of the method for encounter 2 and following}) and apply the technique to a whistler wave packet from Encounter 2 (Section \ref{application}). 


\section{Data and method descriptions}
\label{Method}
%
%
%

\subsection{Data}
\label{data}
The payload of PSP includes a Search-coil Magnetometer (SCM) \cite{jannet_measurement_2021} that measures the 3 Hz to 1 MHz fluctuations of up to three of the orthogonal $(\textbf{u},\textbf{v},\textbf{w})$ components of the magnetic field. The SCM has three low-frequency (LF) windings, one for each component, that cover frequencies from 3 Hz to 20 kHz, and one medium-frequency (MF) winding on the $\mathbf{u}$ component that measures from 1 kHz to 1 MHz; the LF $\mathbf{u}$ winding is the component that became unusable after March 2019. After more than three years of operation, the SCM has revealed a multitude of different wave phenomena in the solar wind, reviewed by \citeA{dudok_de_wit_first_2022}. Among them are whistler waves occurring in the solar wind over a wide range of heliocentric distances.

Complementing the SCM aboard PSP is an Electric Field Instrument (EFI) that measures two components of the electric field from DC to 1 MHz. The EFI uses the PSP spacecraft reference frame, $(\textbf{X},\textbf{Y},\textbf{Z})$, which is different from the $(\textbf{u},\textbf{v},\textbf{w})$ SCM reference frame \cite{malaspina_digital_2016}. A description of these various reference frames as well as the rotation matrix for transforming from the SCM frame to the spacecraft frame is presented in \ref{SC_SCM}. The four electric PSP EFI antennas are located in the plane of the spacecraft heat shield, which is the ($\mathbf{X}$,$\mathbf{Y}$) plane of the spacecraft coordinate system with the $\mathbf{Z}$ axis directed toward the Sun. These antennas, therefore, allow the measurement of the $\mathbf{X}$ and $\mathbf{Y}$ components of the electric field.

Both the SCM and EFI are a part of the PSP FIELDS suite \cite{bale_fields_2016,malaspina_digital_2016}. The data products from FIELDS for Encounter 1 include continuous waveforms with sampling rate of 292.97 $\si{\hertz}$ in the vicinity of perihelia (146 $\si{\hertz}$ and 73 $\si{\hertz}$ at larger distances from the Sun) and 3.5 $\si{\second}$ burst waveform intervals with 150 $\si{\kilo \hertz}$ sampling rate (up to a few dozen a day in the vicinity of perihelia). There are also continuous cross-spectra (every 27.96 $\si{\second}$) that enable the polarization properties of whistler waves to be determined over the frequency range of 23 to 4541 \si{\hertz} in 54 logarithmically spaced frequency channels. Finally, there are Band-Pass Filtered data (BPF) that provide the amplitude of the magnetic field every 0.87 \si{\second} with a lower frequency resolution.

In this study, we use the survey mode waveforms (292.97 $\si{\hertz}$) and the burst waveforms (150 $\si{\kHz}$) of the magnetic and electric fields from the SCM and EFI. We also use data from the PSP DC fluxgate magnetometer (MAG), that measures three components of the magnetic field in spacecraft coordinates, to estimate the background magnetic field. The background plasma density and the solar wind speed are obtained from measurements made by the SWEAP Solar Probe Cup, SPAN-C \cite{kasper_solar_2016,case_solar_2020}.

\subsection{The reconstruction technique}
\label{The reconstruction technique}
With three components of the fluctuating magnetic field the ellipticity of the wave can be obtained by the analysis of the spectral matrices \cite{means_use_1972,santolik_singular_2003,taubenschuss_wave_2019}. The WNA can be determined with these matrices or by a minimum variance analysis \cite{sonnerup_magnetopause_1967,sonnerup_minimum_1998,paschmann_analysis_1998}. These methods give the propagation angle with an 
ambiguity of $\pm 180 ^\circ$ which is removed by calculating the radial direction of propagation (sunward or anti-sunward). The latter is determined by calculating the sign of the Z component ($S_{\rm Z}$, of the spacecraft reference frame) of the Poynting flux. \\ 
The three components of the magnetic field and the two components of the electric field can also be used with the equation $\vec{E}{\rm w}(\omega,t) \cdot \vec{B}{\rm w}(\omega,t)=0$ to find the missing component of the electric field and thus determine the Poynting flux completely. \\

 If one of the three components of magnetic field perturbations is not measured, direct estimation of wave polarization parameters is not possible. However, if two magnetic components and two electric field components are  and geometrically independent, as in the case of the SCM and EFI, it can be possible to reconstruct the missing third component of the fluctuating magnetic field; this requires knowledge of the  wave dispersion and polarization properties over the range of observed perturbation frequencies. Whistler waves, as an electromagnetic plasma mode with a well-defined right-handed polarization in the frequency range from $f_{\rm lh}$ to $f_{\rm ce}$, are a good candidate for reconstruction.

Using the cold plasma dispersion relation for whistler waves and the low-frequency and high-density limits ($\omega/\omega_{\rm ce}\ll1$, $\omega_{\rm pe}^2\gg\omega_{\rm ce}^2$ where $\omega_{\rm pe}$ is the local electron plasma frequency, $\omega$ the wave frequency in the plasma frame and $\omega_{\rm ce}=2\pi f_{\rm ce}$), one can show that (see details in Appendix \ref{Estimation of the parallel electric field component in the spacecraft frame}):  
  \begin{equation}
        \frac{|E^{\rm SC}\rm w_{||}| }{|E{\rm w}|}	\leq  (\frac{V_{\rm SW\perp}}{V_{\varphi}} + (\frac{\omega}{\omega_{\rm ce}}) \tan{\theta} )
   \label{Eparr_final}
 \end{equation} 
where $|E^{\rm SC}\rm w_{||}|$ is the modulus of the electric field component along the background magnetic field $\vec{B}_{\rm 0}$ in the spacecraft frame. In the following, quantities with the $^{\rm SC}$ superscript are in the spacecraft frame while quantities in the plasma frame (i.e., taking into account the Lorentz transformations \cite{feynman_feynman_1964}) are noted without superscript. Subscripts are used to give the reference frames, $(\mathbf{X},\mathbf{Y},\mathbf{Z})$ corresponding to the spacecraft, $(\mathbf{u},\mathbf{v},\mathbf{w})$ corresponding to the SCM and $(\parallel,\perp)$ corresponding to the background magnetic field. $V_{\rm SW\perp}$ is the measured perpendicular solar wind speed and $V_{\varphi}$ is the wave phase speed. 
We note that $|E^{\rm SC}\rm w_{||}|/|E{\rm w}|$ is small if $\omega \ll \omega_{\rm ce}$, $\tan \theta \leq 1$ and $V_{\rm SW\perp}/V_{\varphi} \ll 1$. In this case, we can make the following approximation:


  \begin{equation}
E^{\rm SC}{\rm w}_{\rm X} B_{\rm 0X} + E^{\rm SC}{\rm w}_{\rm Y} B_{\rm 0Y} + E^{\rm SC}{\rm w}_{\rm Z} B_{\rm 0Z} = E^{\rm SC}{\rm w}_{||} \simeq 0
   \label{equation1_prime}
 \end{equation} 

The validity of this approximation and its effect on reconstruction is discussed in detail in Section \ref{Approximations}. Equation \ref{equation1_prime} enables a reconstruction of the third component ($E^{\rm SC re}{\rm w}_{\rm Z}$) of the electric field from the measured values $E^{\rm SC}{\rm w}_{\rm X}$ and $E^{\rm SC}{\rm w}_{\rm Y}$. We note that there is a problem when the $B_{\rm 0Z}$ component is close to 0 but this represents only a minority of cases in the PSP measurements. \\
The reconstructed electric field $\vec{E}^{\rm SC re}{\rm w}$ (in the spacecraft reference frame) can then be expressed in the SCM reference frame and used to reconstruct the third component of wave magnetic field $B^{\rm SC re}{\rm w}_{\rm u}$. For this purpose, we use the two measured components of the SCM ($B^{\rm SC}{\rm w}_{\rm v}$ and $B^{\rm SC}{\rm w}_{\rm w}$) and the electromagnetic wave equation in which the only unknown is $B^{\rm SC re}{\rm w}_{\rm u}$:

\begin{equation}
        \vec{E}^{\rm SC re}{\rm w}(\omega, t)\cdot \vec{B}^{\rm SC}{\rm w}(\omega,t) \simeq 0
 \label{equation2_final}
 \end{equation}

The right-hand side of Equation \ref{equation2_final} is not explicitly zero because of the approximation made in Equation \ref{equation1_prime}. $(\omega, t)$ represents the time averaged Fourier transform. Equation \ref{equation2_final} is solved in the Fourier frequency domain to take into account the whistler dispersion relation and the dependence of the estimation error on frequency. The waveform $B^{\rm SC re}{\rm w}_{\rm u}$ is then obtained from the inverse Fourier transform. As explained in Appendix \ref{Estimation of the parallel electric field component in the spacecraft frame}, because $V_{\rm SW} \ll c$ (where c is the speed of light) we can safely consider that $\vec{B}^{\rm SC}{\rm w} = \vec{B}{\rm w}$. 

 It should be noted that the form of Equations \ref{equation1_prime} and \ref{equation2_final} shows that the method is independent of the chosen effective length of the antennas. In the following, we therefore take an effective length of 3.5 $\si{\meter}$, even though this length depends on the frequency and on the propagation direction \cite{karbashewski_whistler_2023}. 


\color{black}

\subsection{Validity of the approximation}
\label{Approximations}

\begin{figure}
\includegraphics[width=100ex]{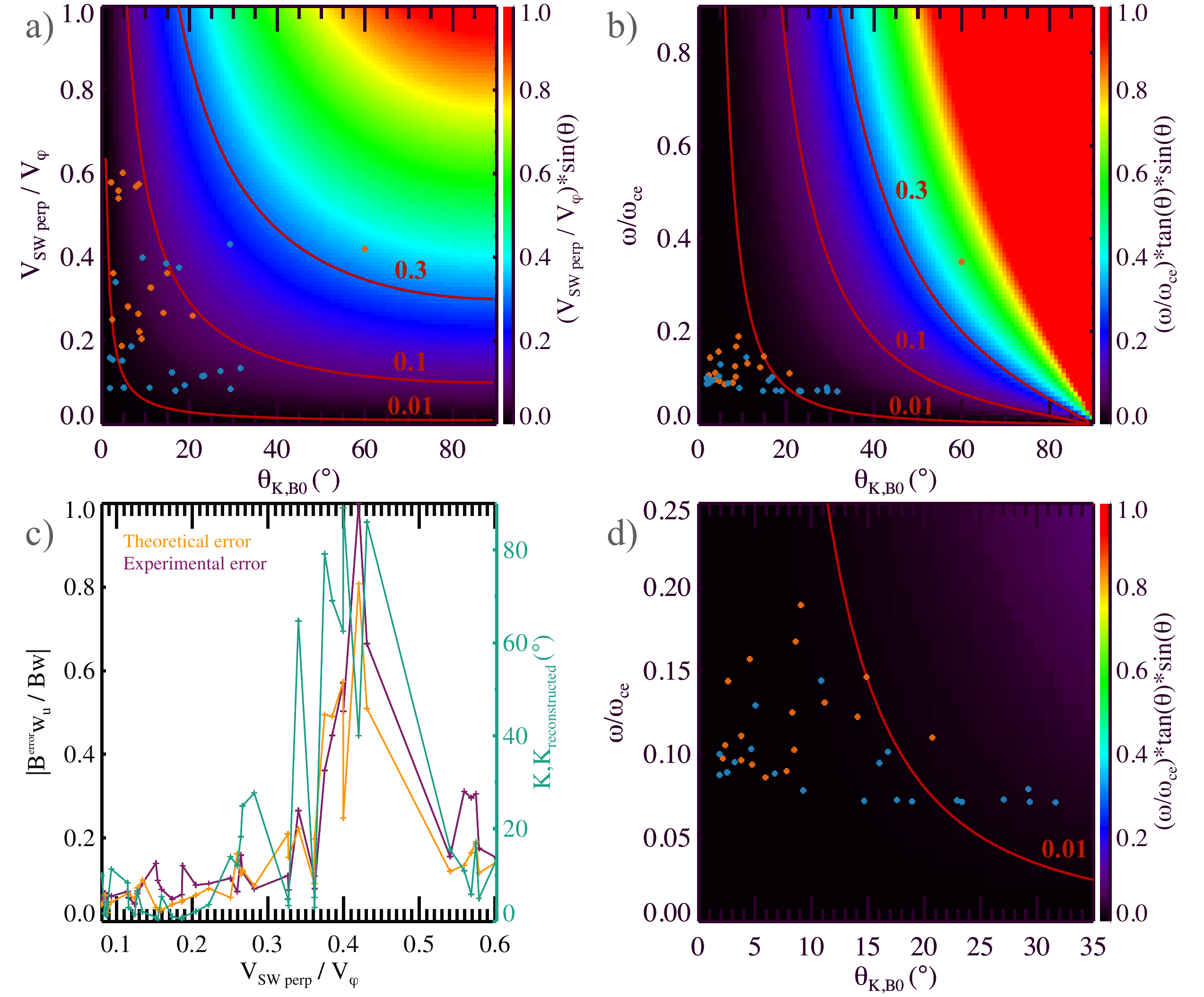}
\caption{
(a) - Parameters of whistler waves in the ($\theta,\frac{V_{\rm SW\perp}}{V_{\varphi}})$ plane observed by PSP during encounter 1 (see text). Anti-sunward (resp., sunward) whistler waves are represented by blue (resp., orange) dots. The error due to the  term $\frac{V_{\rm SW\perp}}{V_{\varphi}} \sin{\theta}$ is indicated by the background color. (b) - parameters of whistler waves in the ($\theta,\omega / \omega_{\rm ce}$) plane. The error due to the  term ($\frac{\omega}{\omega_{\rm ce}}) \tan{\theta} \sin{\theta}$ is indicated by the background color. (c) - theoretical and experimental errors of the reconstruction technique, as well as the angular deviation between the measured and reconstructed wave vector as a function of $\frac{V_{\rm SW\perp}}{V_{\varphi}}$. (d) - zoom on the on the lower left side of the panel (b). In panels a),c) and d) $\omega$ is the frequency in the plasma frame.}
\label{error_final}
\label{error_theo_exp}
\end{figure}

The main approximation of this method is therefore to consider $E^{\rm SC}{\rm w}_{||}=0$. An upper value of $|E^{\rm SC}{\rm w}_{||}|/|E{\rm w}|$ is given by Equation \ref{Eparr_final}. We can distinguish two sources of error. \\
The first source of error comes from the fact that the measured electric field is different from the electric field in the plasma frame.
To obtain the electric field in the plasma frame, one needs to carry out the Lorentz transformations \cite{feynman_feynman_1964}, which is not directly possible with only two components of the magnetic field. This error can be expressed as $\frac{V_{\rm SW\perp}}{V_{\varphi}}$ and can be significant if the solar wind velocity and background magnetic field are not aligned and if the phase velocity is low. \\
The second source of error comes from the parallel component of the electric field in the plasma frame, which is not zero if the WNA is not equal to zero. This error can be expressed as $(\frac{\omega}{\omega_{\rm ce}}) \tan{\theta}$ and can be significant if the wave is oblique or if $\frac{\omega}{\omega_{\rm ce}}$ is large. 

 The propagation of these errors gives an error on the reconstructed magnetic field, whose amplitude can be approximated as follows (see details in appendix \ref{Propagation of the error}):

\begin{equation}
    ( |B^{\rm error}{\rm w}_{\rm u}  / B{\rm w}| ) \lesssim \sqrt{( \frac{V_{\rm SW\perp}}{V_{\varphi}} \sin{\theta})^2 + (  (\frac{\omega}{\omega_{\rm ce}}) \tan{\theta} \sin{\theta} )^2 }  \frac{1}{ \sin{\theta_{\Vec{B_{\rm 0}}, \Vec{u}}}} 
    \label{propagation_erreur}
\end{equation}

We note a multiplication by $\sin{\theta}$ (compared to Equation \ref{error_final}), which can be explained by the fact that the error is on the parallel component of the electric field, which is multiplied (Equation \ref{equation2_final}) by the parallel component of the magnetic field. Finally, the term $\sin{\theta_{\Vec{B_{\rm 0}}, \Vec{u}}}$ comes from the fact that we can't reconstruct the parallel component of the fields correctly since we are making the assumption $E^{\rm SC}{\rm w}_{||} = 0$. Thus, there is a problem when $\Vec{B_{\rm 0}}$ and $\Vec{u}$ are aligned. 

In Figure \ref{error_final}a we represent $\frac{V_{\rm SW\perp}}{V_{\varphi}} \sin{\theta}$ in the ($\theta,\frac{V_{\rm SW\perp}}{V_{\varphi}})$ plane. The whistler wave packets measured during Parker Solar Probe's first encounter using burst mode are superimposed on this panel (50 wave packets were detected). We also add the characteristics of a wave packet from continuous waveforms at 292.97 $\si{\hertz}$ (point with $\theta \simeq 60^\circ$). This wave packet is studied in detail in Section \ref{case3}. Whistler detection and characterization methods are described in \citeA{kretzschmar_whistler_2021}. We note that the vast majority of points (96\%) are below the 10 \% error line. \\
In Figure \ref{error_final}b we represent $\frac{\omega}{\omega_{\rm ce}} \tan{\theta} \sin{\theta}$ in the $(\theta,\frac{\omega}{\omega_{\rm ce}})$ plane. The measured whistler characteristics are also superimposed on this panel (see Figure \ref{error_final}a). Note that the vast majority (99\%) of points are below the 10\% error line and 92\% of points are below the 0.1\% line.
We can note that the frequency in the plasma frame of the sunward waves is generally greater than that of the anti-sunward waves (Figure \ref{error_final}d). This is a good indication of validity of the generation mechanism proposed by \citeA{karbashewski_whistler_2023}. Note that in this study we are only interested in the clearly sunward or anti-sunward cases, and do not consider the counter-streaming cases \cite{karbashewski_whistler_2023}.


The final theoretical error (given by Equation \ref{propagation_erreur}) is plotted as a function of $\frac{V_{\rm SW\perp}}{V_{\varphi}}$ on Figure \ref{error_final}c using the characteristics of the measured whistler wave packets. This error is compared with the experimental error defined as $mean(|B^{\rm re}{\rm w}_{\rm u} - B{\rm w}_{\rm u}|/max(B{\rm w}_{\rm u}))$. There is a good agreement between these two curves which is a good indication that the error is well estimated by Equation \ref{propagation_erreur}. We note that theoretical and experimental errors tend to increase with $\frac{V_{\rm SW\perp}}{V_{\varphi}}$. We also observe a significant error for cases with $0.35<\frac{V_{\rm SW\perp}}{V_{\varphi}}<0.5$ which is due to the fact that for the majority of these cases $\vec{B_{\rm 0}}$ and $\Vec{u}$ were almost aligned. The decrease in error for cases verifying $\frac{V_{\rm SW\perp}}{V_{\varphi}} > 0.5$ is due to the fact that their WNAs are close to 0$^\circ$. In addition, when the theoretical and experimental errors are large, there is a large discrepancy between the measured and reconstructed wave vectors. \\

For $78\%$ of cases, the theoretical and experimental errors are less than 0.2. For these cases, the error between the measured and reconstructed wave vectors is always less than $30^\circ$ with an average value of $6^\circ$. For 100\% of these cases, the reconstructed direction of propagation (sunward or anti-sunward) was found to be correct. Moreover, the averaged reconstructed ellipticity is 0.80, compared with 0.85 for the measured wave packets. \\
For cases where the error is greater than 0.2, large discrepancies are found between the measured and reconstructed wave vectors (up to $90^\circ$), and the sunward or anti-sunward propagation direction is found in only 65 $\%$ of the cases. For these cases, the averaged reconstructed ellipticity is 0.46.  

We can therefore conclude that in about 80\% of the burst mode data from the first Parker Solar Probe encounter, the reconstruction method is applicable. When the theoretical error (given by Equation \ref{propagation_erreur}) is below 0.2, the technique allows to find the direction of propagation (sunward or anti-sunward) in 100 $\%$ of cases and the error on the WNA is on average 6$^\circ$. When the theoretical error is bigger than 0.2, the technique is not applicable and this is mainly due to a high $\frac{V_{\rm SW\perp}}{V_{\varphi}}$ or a low $\theta_{\Vec{B_{\rm 0}}, \Vec{u}}$. This last source of errors can be easily checked. On the other hand,  $\theta$, $V_{\varphi}$ and $\omega$ are no longer directly accessible after March 2019. The applicability of this method after this date is therefore discussed in detail Section \ref{Discussion on the applicability of the method for encounter 2 and following}. 

Finally, as mentioned above, Equation \ref{Eparr_final} (and therefore Equation \ref{propagation_erreur}) are based on the high density hypothesis $\omega_{\rm pe}^2\gg \omega_{\rm ce}^2$. 
Extrapolation of HELIOS data (between 0.3 and 1 AU, \cite{bale_fields_2016}) shows that the expected $\omega_{\rm pe}^2/\omega_{\rm ce}^2$ is around 150 at 10 $R_{\odot}$. This ratio increases with distance, indicating that this assumption should be valid for all Parker Solar probe encounters.

The results of the reconstruction and its accuracy are illustrated below on three examples from Encounter 1 when all three (\textbf{u},\textbf{v},\textbf{w}) components of the SCM were available. Two examples demonstrate the regularly observed by PSP whistler wave characteristics: Case 1 and Case 2. Case 3 is atypical because of its frequency, its phase speed, and its WNA and illustrates what happens when one is out of the applicability range of the method. We provide the results of polarization analysis (the radial component of Poynting flux, WNA, ellipticity) and the power spectral density to compare with the values obtained by making use of the reconstructed $B^{\rm re}{\rm w}_{\rm u}$ magnetic field component. For these 3 examples, we first present the case using the actual measurements (Figures \ref{fig:case1},\ref{fig:case2} and \ref{fig:case3}) and then compare them with the results of the reconstruction (Figures \ref{fig:case1reconstruction},\ref{fig:case2reconstruction} and \ref{fig:case3reconstruction}). 

\color{black}
\section{Application of the method}
\subsection{Case 1: November 3, 2018, 10:33:31.0-10:33:34.5 UT}
\label{Case1}

\begin{figure}
 \noindent\includegraphics[width=100ex]{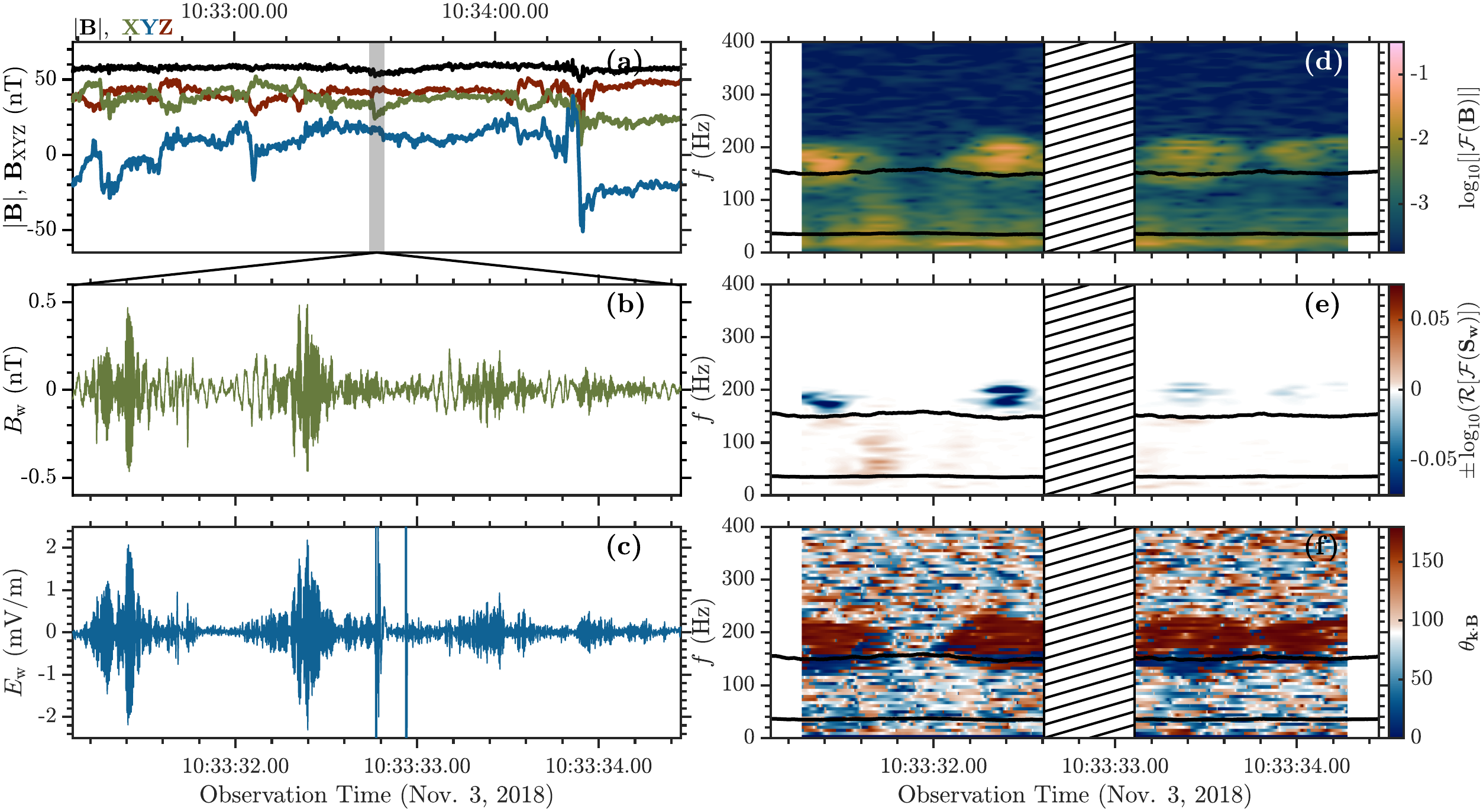}
\caption{Whistler wave packets recorded by PSP on November 3, 2018. (a) - background magnetic field in the spacecraft coordinates over a short window around the burst. (b) - burst waveform of the Y magnetic field component, $B{\rm w}_{\rm Y}$. (c) - burst waveform of the Y electric field component, $E{\rm w}_{\rm Y}$. (d) - spectrogram of the magnetic field burst waveforms. (e) - Z component of the Poynting flux. (f) - WNA $\theta_{\mathbf{k\cdot B}}$, ranging from $0^\circ$ to $180^\circ$ and indicating parallel (below 90$^\circ$) and anti-parallel propagation (above 90$^\circ$), respectively. The lower and upper solid lines in panels (d)-(f) indicate $f_{\rm lh}$ and $0.1f_{\rm ce}$, respectively.
For panels (d) to (f) the frequency is shown in the spacecraft frame. 
\label{fig:case1}
}
\end{figure}

Figure \ref{fig:case1} presents two typical anti-sunward propagating whistler wave packets, recorded at the heliocentric distance of $\sim41$ ($R_{\odot}$) in the FIELDS burst mode (150 $\si{\kilo \hertz}$) at 10:33:31.0 UTC, November 3, 2018. This figure is adapted from Figure 9 of \citeA{karbashewski_whistler_2023}. The polarization parameters are directly evaluated using measurements of the three magnetic field components and the missing component of the electric field is estimated from $\vec{E}{\rm w}(\omega,t) \cdot \vec{B}{\rm w}(\omega,t)=0$ (Figures \ref{fig:case1}c and \ref{fig:case1}e). This allows us to estimate the Poynting flux vector. The Poynting flux reveals that the whistler waves are propagating from the Sun almost field-aligned, in the opposite direction to the background magnetic field (Figures \ref{fig:case1}e and \ref{fig:case1}f). These wave packets are not associated with any significant perturbation of the background magnetic field (Figure \ref{fig:case1}a), which is regular for anti-sunward propagation \cite{karbashewski_whistler_2023}. The observed wave and plasma parameters are typical of the young solar wind: the background magnetic field magnitude is 55 nT; the plasma density is $\sim290$ $cm^3$; $(\omega_{\rm pe}/\omega_{\rm ce})^2 \sim$ 10 000; $\omega/\omega_{\rm ce} \sim 0.1$ (in the plasma frame); the bulk radial plasma velocity is 310 $\si{\kilo \meter / \second}$; the wave amplitudes reache 0.5 nT. The observed WNAs of the packets are below 20$^\circ$ (Figure \ref{fig:case1}f). The wave packets propagating anti-sunward have the wave frequency downshifted from the range 130-200 Hz (in the spacecraft frame) to 110-180 Hz (0.08-0.13 $f_{\rm ce}$) in the plasma frame. The solar wind velocity perpendicular to the magnetic field $V_{\rm SW \perp}$ is about 134 $\si{\kilo \meter} / \si{\second} $. Finally, the phase velocity is $V_{\rm \varphi}  \sim 893 \si{\kilo \meter} / \si{\second}$ and the component to be reconstructed for this test event satisfies the condition of being nearly perpendicular to the magnetic field $\sin(\theta_{\vec{u},\vec{B_0}}) \sim 0.99$. 

\begin{figure}[H]
\includegraphics[width=100ex]{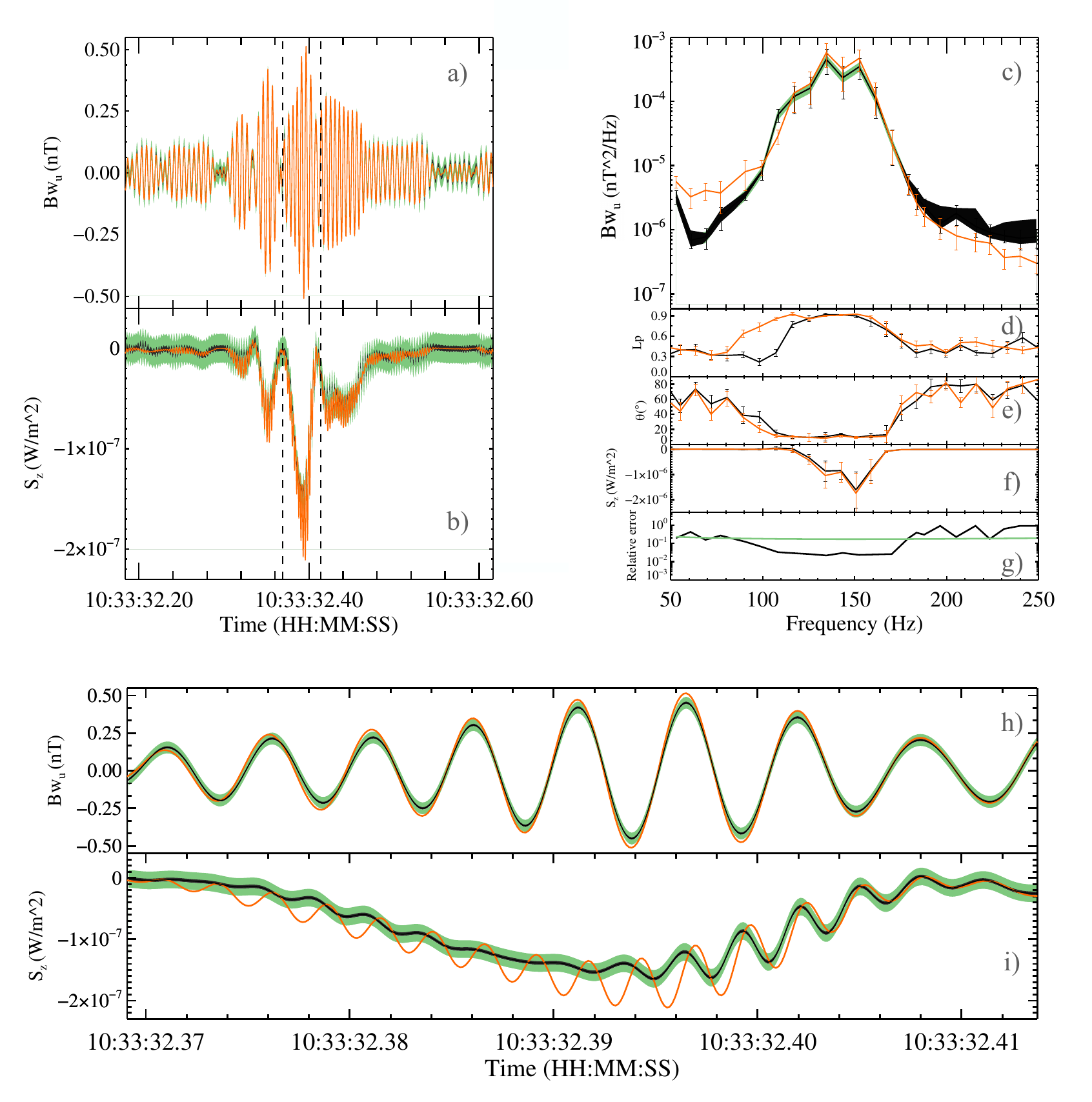}
\caption{Comparison between the whistler wave packet recorded by PSP on November 3, 2018 and its reconstruction. The original data are in orange and the reconstructed ones are in black. The error bars are detailed in the text.
(a) - burst waveform of the u magnetic field component. (b) - Z component of the Poynting flux. 
The panels (c) to (g) show the frequency in the plasma frame.  
(c) - power spectral density of the u component. (d) - ellipticity. (e) - WNA $\theta_{\mathbf{k\cdot B}}$. (f) - Z component of the Poynting flux. (g) - theoretical relative error in black and using a proxy of the angle of propagation in green.  
Panels (h) and (i) show a zoom on the period between the vertical dotted lines in panels (a) and (b).  
}
\label{fig:case1reconstruction}
\end{figure}

The results of wave polarization reconstruction on the first wave packet are shown in Figure \ref{fig:case1reconstruction}.
The orange curves represent the original data, while the black and green curves are obtained using the reconstructed $B^{\rm re}{\rm w}_{\rm u}$ component. For Figures \ref{fig:case1reconstruction}a, \ref{fig:case1reconstruction}b, \ref{fig:case1reconstruction}h, and \ref{fig:case1reconstruction}i the thickness of the green and black lines corresponds to the estimation of the theoretical error of the technique (Equation \ref{propagation_erreur}).
This relative error is calculated using the typical frequency (in the plasma frame) and angle of propagation of the wave (black) and a proxy of the propagation angle in green. This proxy is estimated using the theoretical ratio $E_{\rm w||}/E_{\rm w}$ in the plasma frame (Equation \ref{E_plasma_frame}). For a given value of $\omega/\omega_{\rm ce}$ the curve $E_{\rm w||}/E_{\rm w}$ as a function of $\theta$ has a plateau shape. The proxy of the WNA represents the mean value of $\theta$ on this plateau. For Figure \ref{fig:case1reconstruction}c the theoretical errors (associated with each frequency and angle of propagation) are in black and using a proxy of the angle of propagation in green.  These theoretical relative errors are shown in Figure\ref{fig:case1reconstruction}g  and are limited to 1. Figures \ref{fig:case1reconstruction}c to \ref{fig:case1reconstruction}f the error bars correspond to the statistical errors of the computation of spectral matrices. 
The case satisfies very well the applicability parameters for the reconstruction: $|B^{\rm error}{\rm w}_{\rm u}  / B{\rm w}|$ is in the range 0.02-0.05 for the entire frequency range of the whistler activity (Figure \ref{fig:case1reconstruction}g). The results obtained from the reconstructed $B^{\rm re}{\rm w}_{\rm u}$ are in very good agreement with the results based on the measured $B{\rm w}_{\rm u}$. Indeed, we can see in Figures \ref{fig:case1reconstruction}a, \ref{fig:case1reconstruction}b, \ref{fig:case1reconstruction}h, and \ref{fig:case1reconstruction}i that there is a very good agreement (phase and amplitude) between the waveforms. The initially measured waveforms are very often contained in the error bars. This shows that the error is estimated adequately. Furthermore, we can see in Figures \ref{fig:case1reconstruction}b, \ref{fig:case1reconstruction}f, and \ref{fig:case1reconstruction}i that the Z component of the Poynting vector is very well reconstructed, allowing the propagation direction to be recovered without ambiguity. 
We can also reconstruct the spectrum (Figure \ref{fig:case1reconstruction}c) in a satisfactory manner, the whistler spectral bump is clearly identified. The measured and reconstructed ellipticity values are greater than 0.6 over the entire frequency range of the wave. The WNA $\theta$ (Figure \ref{fig:case1reconstruction}f) is also in very good agreement with the measurement and the typical error on the frequency range of the wave is of the order of a degree. Finally, the minimum variance analysis gives less than $2^\circ$ deviation between the wave vectors using the measured and reconstructed $B{\rm w}_{\rm u}$ (not shown).


Thus, for this anti-sunward propagating wave packet, which clearly satisfies the applicability parameters for reconstruction, the technique works effectively. Specifically, all important reconstructed parameters are in good agreement with the measurements: amplitude, ellipticity, direction of propagation, and WNA (with an error of less than 2$^\circ$).


\subsection{Case 2: November 3, 2018, 10:19:15.6-10:19:19.4 UT} 
\label{Case 2}
\begin{figure}[H]
\noindent\includegraphics[width=100ex]{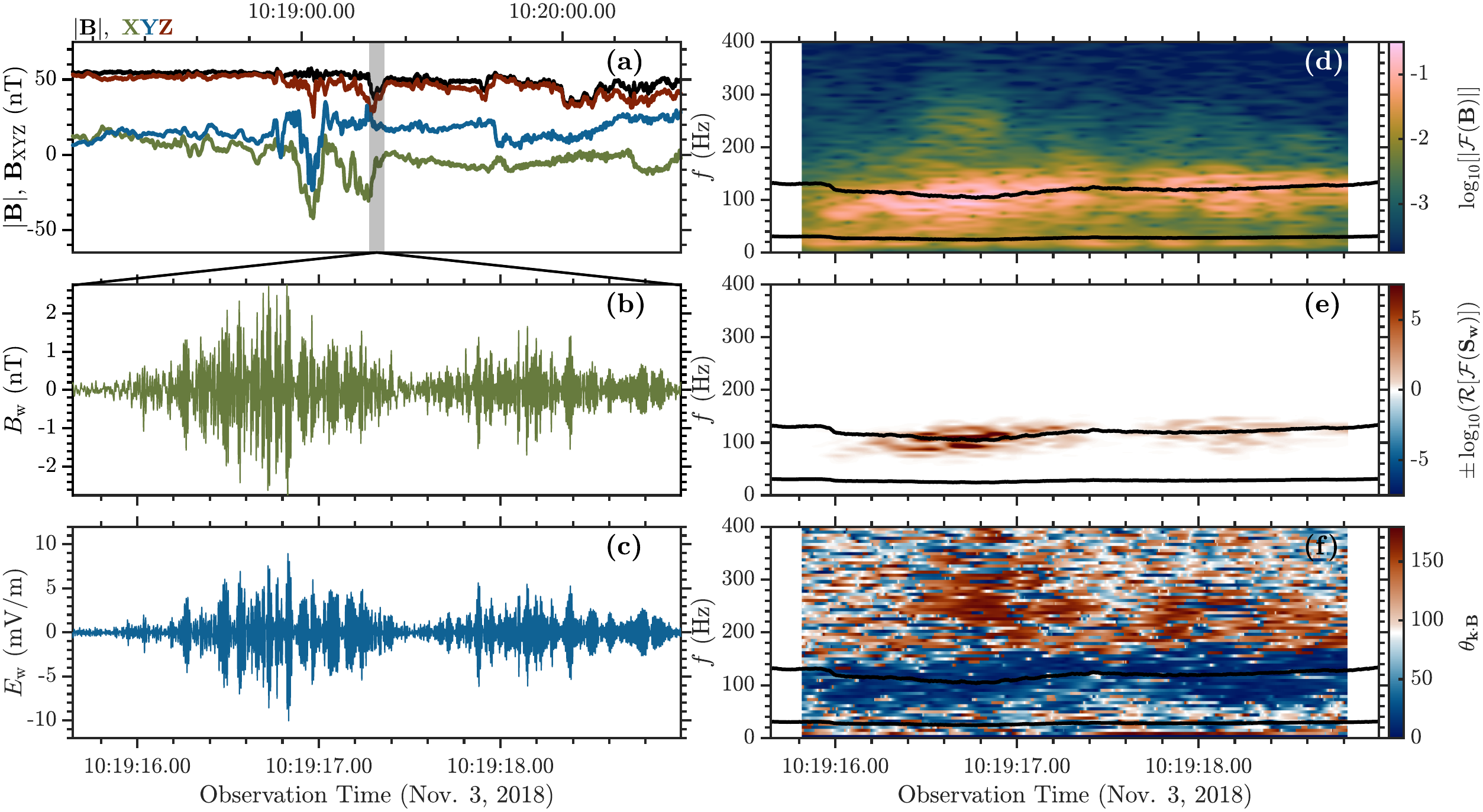}
\caption{Whistler wave packets recorded by PSP on November 3, 2018. (a) - backgroud magnetic field in the spacecraft coordinates over a short window around the burst. (b) - burst waveform of the Y magnetic field component, $B{\rm w}_{\rm Y}$. (c) - burst waveform of the Y electric field component, $E{\rm w}_{\rm Y}$. (d) - spectrogram of the magnetic field burst waveform. (e) - Z  component of the Poynting flux. (f) - WNA $\theta_{\mathbf{k\cdot B}}$, ranging from $0^\circ$ to $180^\circ$ and indicating parallel (below 90$^\circ$) and anti-parallel propagation (above 90$^\circ$), respectively. The lower and upper solid lines in (d)-(f) indicate $f_{\rm lh}$ and $0.1f_{\rm ce}$, respectively.
For panels (d) to (f) the frequency is shown in the spacecraft frame. \label{fig:case2}}
\end{figure}

Figure \ref{fig:case2} presents two typical sunward whistler wave packets, recorded at the heliocentric distance of $\sim41$ ($R_{\odot}$) in the FIELDS burst mode (150 $\si{\kilo \hertz}$) at 10:19:15.57 UTC, November 3, 2018. This figure is adapted from Figure 6 of \citeA{karbashewski_whistler_2023}. The polarization parameters and the missing component of the electric field are evaluated as in Section \ref{Case1}. The Poynting flux shows a sunward propagation, quasi-aligned with the background magnetic field (Figures \ref{fig:case2}e and \ref{fig:case2}f). As we can see in Figure \ref{fig:case2}a, the wave packets are associated with a dip of the background magnetic field of the order of 20 $\%$. This is expected for sunward whistler waves \cite{karbashewski_whistler_2023}. The background magnetic field magnitude is 48 nT; the plasma density is $\sim410$ $cm^3$; $(\omega_{\rm pe}/\omega_{\rm ce})^2 \sim$ 28 900;  $\omega/\omega_{\rm ce} \sim 0.17$ (in the plasma frame); the bulk radial plasma velocity is 310 $\si{\kilo \meter / \second}$; the wave amplitudes reache 2.5 nT. The observed WNAs are below 30$^\circ$ (Figure \ref{fig:case2}f). The wave packets propagating sunward have the wave frequency shifted from the range 60-160 Hz (in the spacecraft frame) to 120-240 \si{\hertz} (0.11-0.22 $f_{\rm ce}$, in the plasma frame). The solar wind velocity perpendicular to the magnetic field $V_{\rm SW \perp}$ is about 194 $\si{\kilo \meter} / \si{\second} $. The phase velocity is $V_{\rm \varphi}  \sim$ 562 $\si{\kilo \meter} / \si{\second}$ and $\sin(\theta_{\vec{u},\vec{B_{\rm 0}}}) \sim 0.76$.

\begin{figure}[H]
\includegraphics[width=100ex]{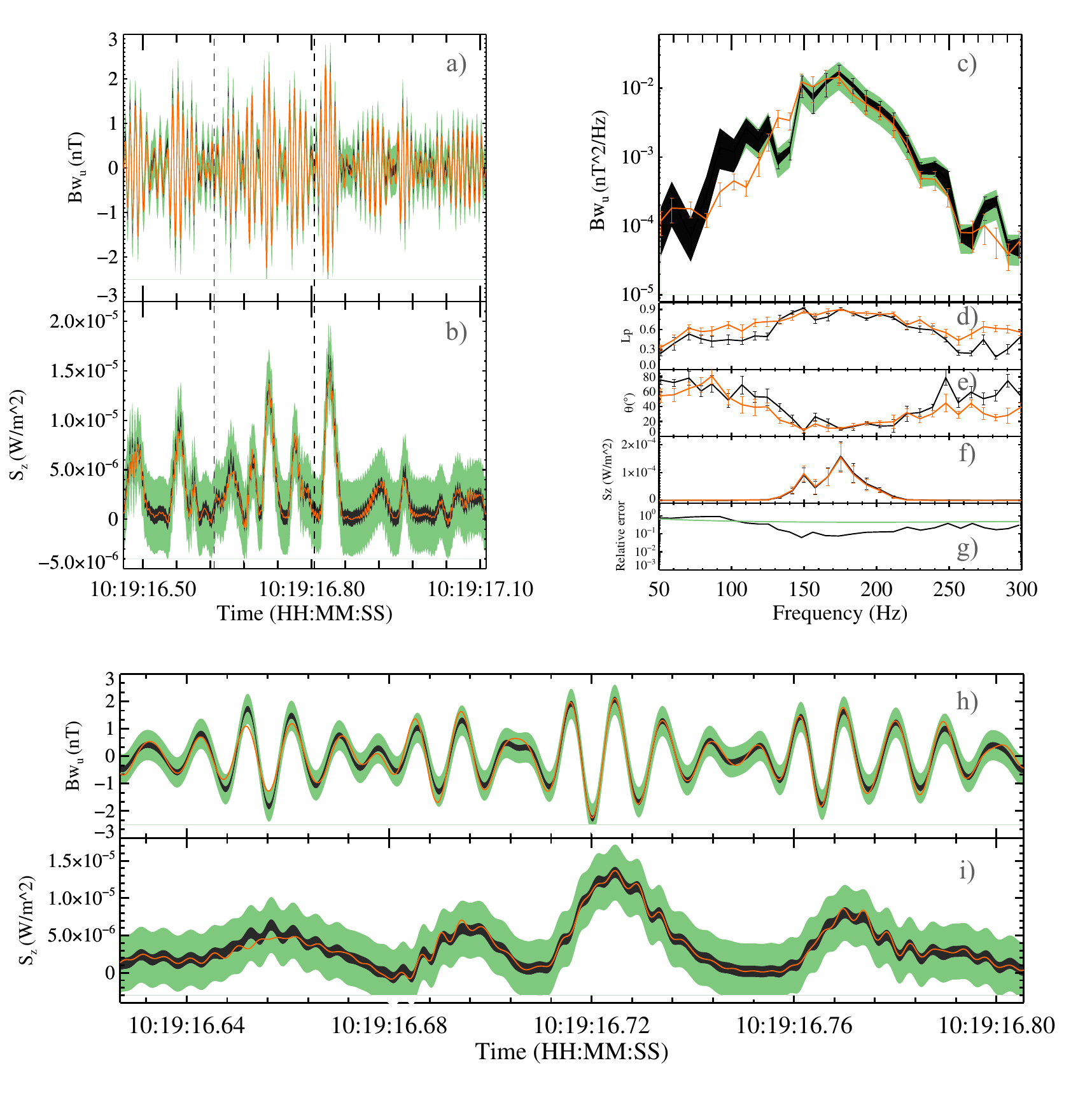}
\caption{Comparison between the whistler wave packet recorded by PSP on November 3, 2018 and its reconstruction. The original data are in orange and the reconstructed ones are in black. The error bars are detailed in Section \ref{Case1}. 
(a) - burst waveform of the u magnetic field component. (b) - Z component of the Poynting flux. 
The panels (c) to (g) show the frequency in the plasma frame.  
(c) - power spectral density of the u component. (d) - ellipticity. (e) - WNA $\theta_{\mathbf{k\cdot B}}$. (f) - Z component of the Poynting flux. (g) - theoretical relative error in black and using a proxy of the angle of propagation in green. Panels (h) and (i) show a zoom on the period between the vertical dotted lines in panels (a) and (b). }
\label{fig:case2reconstruction}
\end{figure}

The results of wave polarization reconstruction of the first whistler wave packet are shown in Figure \ref{fig:case2reconstruction}. The color code is the same as in Figure \ref{fig:case1reconstruction}.
This case satisfies the applicability parameters for the reconstruction: $|B^{\rm error}{\rm w}_{\rm u}/B{\rm w}|$ is in the range 0.06-0.2 for the entire frequency range of the whistler activity (Figure \ref{fig:case2reconstruction}g).
Because of a larger $\omega/\omega_{\rm ce}$ and a lower $V_{\rm \varphi}$ (which is typical for sunward whistlers, see Figure \ref{error_final}) the typical relative errors are about 2 times larger than for Case 1. 
Once again the results obtained from the reconstructed $B^{\rm re}{\rm w}_{\rm u}$ are in very good agreement with the results based on the measured $B{\rm w}_{\rm u}$ and are very similar to those described for Case 1. The reconstructed waveforms are in good agreement with those originally measured (Figures \ref{fig:case2reconstruction}a, \ref{fig:case2reconstruction}b, \ref{fig:case2reconstruction}h and \ref{fig:case2reconstruction}i). 
With the reconstructed data we can find without ambiguity the characteristics of a whistler wave packet propagating anti-sunward (Figures \ref{fig:case2reconstruction}b, \ref{fig:case2reconstruction}d, \ref{fig:case2reconstruction}f, and \ref{fig:case2reconstruction}i). Figure \ref{fig:case2reconstruction}f shows that the error on the propagation angle is of the order of a few degrees over the frequency range of the wave. The minimum variance analysis gives less than $2^\circ$ deviation between the wave vectors using the measured and reconstructed $B{\rm w}_{\rm u}$ (not shown).

Thus, for this sunward propagating wave packet, in the applicability range of the technique, the reconstruction works effectively. Again, all important reconstructed parameters are in good agreement with the measurements: amplitude, ellipticity, direction of propagation, and WNA (with an error of less than 2$^\circ$).

\subsection{Case 3: November 4, 2018, 17:06:47-17:06:51 UT}
\label{case3}
\begin{figure}[H]
\noindent\includegraphics[width=100ex]{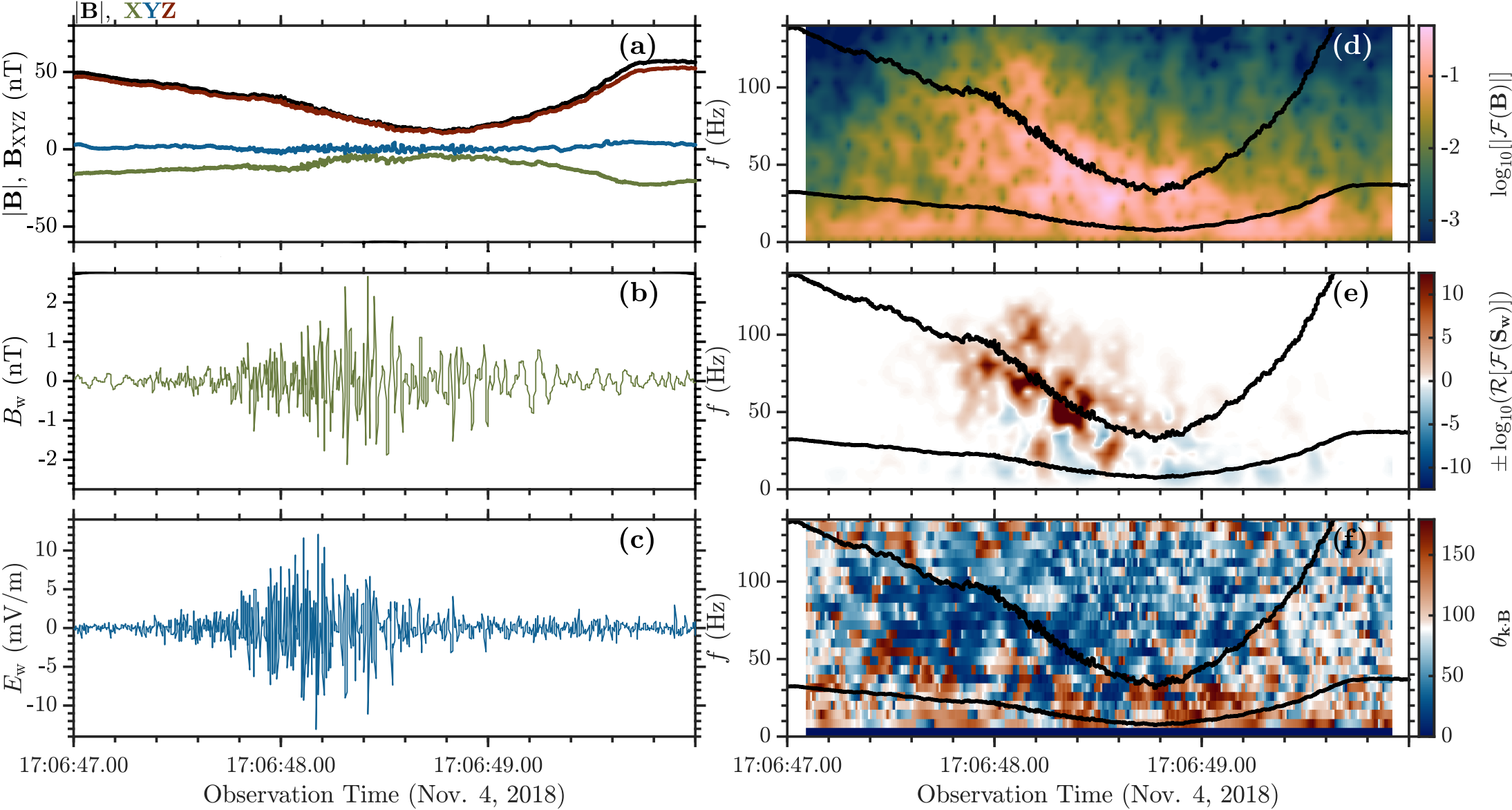}
\caption{Whistler wave packet recorded by PSP on November 4, 2018. (a) - background  magnetic field in spacecraft coordinates. (b) - burst waveform of the Y magnetic field component, $B{\rm w}_{\rm Y}$. (c) - burst waveform of the Y electric field component, $E{\rm w}_{\rm Y}$. (d) - spectrogram of the magnetic field burst waveform. (e) - Z component of the Poynting flux. (f) - WNA $\theta_{\mathbf{k\cdot B}}$, ranging from $0^\circ$ to $180^\circ$ and indicating parallel (below 90$^\circ$) and anti-parallel propagation (above 90$^\circ$), respectively. The lower and upper solid lines in (d)-(f) indicate $f_{\rm lh}$ and $0.1f_{\rm ce}$, respectively.
For panels (d) to (f) the frequency is shown in the spacecraft frame.\label{fig:case3}}
\end{figure}

In Figure \ref{fig:case3} we show the case presented in detail by \citeA{agapitov_sunward-propagating_2020}, recorded in waveforms with a 292.97 $\si{\hertz}$ sampling rate at 17:06:48 UTC, November 4, 2018. We can see a sunward propagating whistler with several oblique WNA sub-packets (Figures \ref{fig:case3}e and \ref{fig:case3}f). The packet is co-located with a local minimum of the background magnetic field magnitude located at a switchback boundary. The minimum $|B_{\rm 0}|$ value is $\sim15$ nT with the ambient magnetic field magnitude of $\sim 70$ nT (Figure \ref{fig:case3}a). The corresponding enhancement of $(\omega_{\rm pe}/\omega_{\rm ce})^2$ (up to 250 000 with the background value of $\sim$ 10 000) inside the magnetic dip causes an unusually large Doppler shift. The wave frequency in the plasma frame is between 0.2 and 0.45 of the local $f_{\rm ce}$. The solar wind velocity perpendicular to the magnetic field $V_{\rm SW \perp}$ is about 160 $\si{\kilo \meter} / \si{\second} $. The phase velocity is $V_{\rm \varphi}  \sim$ 439 $\si{\kilo \meter} / \si{\second}$ and $\sin(\theta_{\vec{u},\vec{B_{\rm 0}}}) \sim 0.56$.

\begin{figure}[H]
\includegraphics[width=100ex]{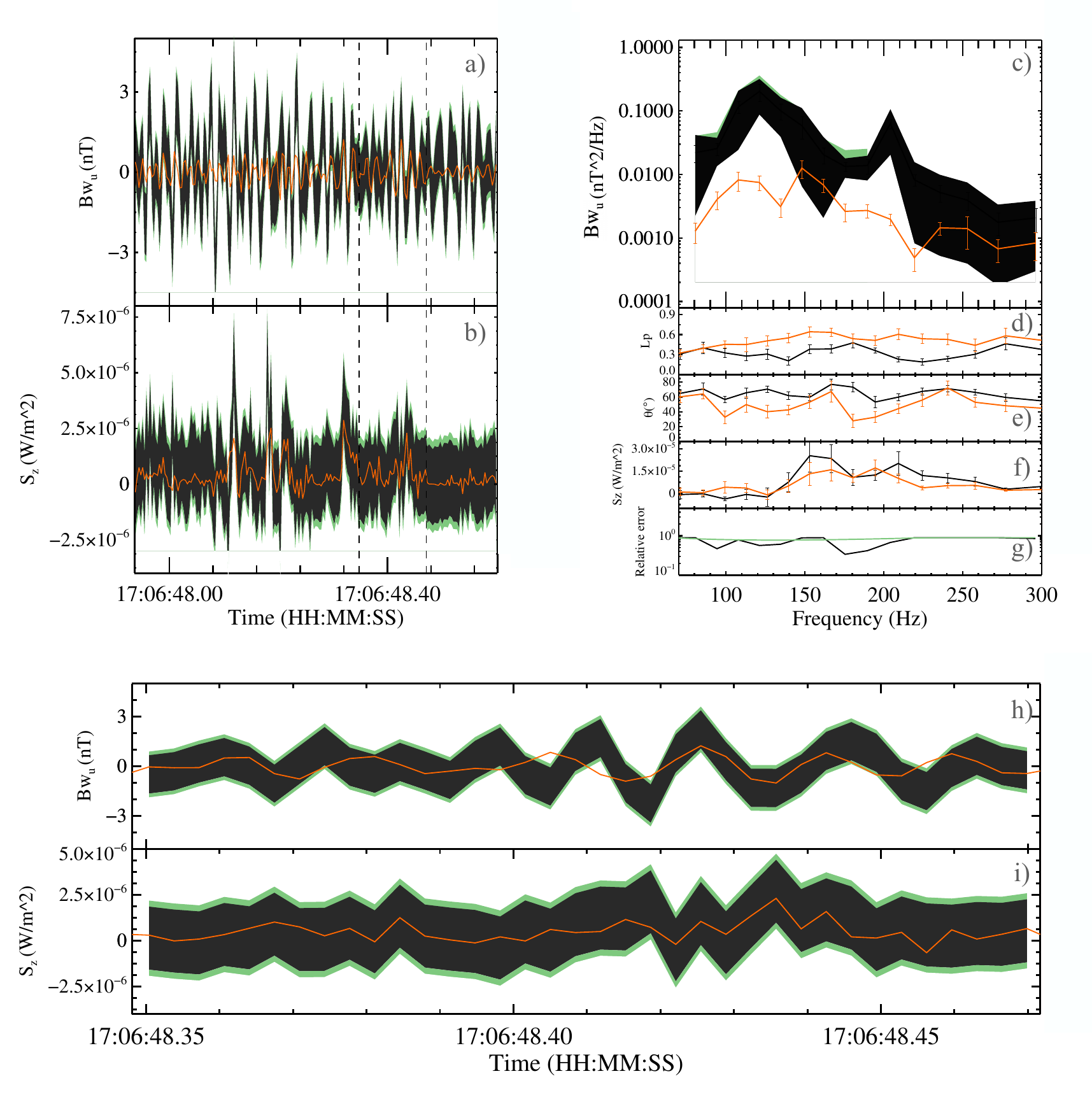}
\caption{Comparison between the whistler wave packet recorded by PSP on November 4, 2018 and its reconstruction. The original data are in orange and the reconstructed ones are in black. The error bars are detailed in Section \ref{Case1}. 
(a) - burst waveform of the u magnetic field component. (b) - Z component of the Poynting flux. The panels (c) to (g) show the frequency in the plasma frame.  
(c) - power spectral density of the u component. (d) - ellipticity. (e) - WNA $\theta_{\mathbf{k\cdot B}}$. (f) - Z component of the Poynting flux. (g) - theoretical relative error in black and using a proxy of the angle of propagation in green.  
Panels (h) and (i) show a zoom on the period between the vertical dotted lines in panels (a) and (b).}
\label{fig:case3reconstruction}
\end{figure}

The results of wave polarization reconstruction are shown in Figure \ref{fig:case3reconstruction}. The color code is the same as in Figures \ref{fig:case1reconstruction} and \ref{fig:case2reconstruction} (explained in Section \ref{Case1}). This is a difficult case for reconstruction. Indeed, as explained above the wave contains several oblique sub-packets (up to $80^\circ$, Figure \ref{fig:case3}f), and the main angle of propagation can be as oblique as $70^\circ$ (Figure \ref{fig:case3reconstruction}e). Moreover, we have spectral energy content up to 120 $\si{Hz}$ (in the spacecraft frame), therefore close to the Nyquist frequency (Figures \ref{fig:case3}d and \ref{fig:case3}e). 
Finally, the main issues are that the wave frequency in the plasma frame is about 0.35 $f_{\rm ce}$ and can be up to 0.45 $f_{\rm ce}$,
$\frac{V_{\rm SW \perp}}{V_{\rm \varphi}}$ is about 0.35 and that $\sin(\theta_{\vec{u},\vec{B_{\rm 0}}}) \sim 0.56$.
Therefore, taking into account the obliquity of the wave, the theoretical relative error is important: $|B^{\rm error}{\rm w}_{\rm u}/B{\rm w}|$ is about 0.5 and can be greater than 1 (Figure \ref{fig:case3reconstruction}g). 
The results obtained from the reconstructed $B^{\rm re}{\rm w}_{\rm u}$ are not in good agreement with the results based on the measured $B{\rm w}_{\rm u}$. As we can see on Figures \ref{fig:case3reconstruction}a , \ref{fig:case3reconstruction}b, \ref{fig:case3reconstruction}h and \ref{fig:case3reconstruction}i the reconstructed waveforms do not approach the initial waveforms well. Important overestimation of the amplitude (about 3 times) is noted in the reconstructed $B^{\rm re}{\rm w}_{\rm u}$ component. The Poynting flux is not perfectly recovered but the sunward direction of propagation is still clear (Figures \ref{fig:case3reconstruction}b, \ref{fig:case3reconstruction}f, and \ref{fig:case3reconstruction}i). The reconstructed spectrum is about an order of magnitude larger than the measured one (Figure \ref{fig:case3reconstruction}c). The reconstructed ellipticity is lower than 0.6 on all frequencies of the waves, which does not allow us to find the classical characteristics of a whistler wave. The propagation angle is wrong by $40^\circ$ for some frequencies, which can lead to a misinterpretation of the effect of the wave on the electrons.


Thus, for this case out of the applicability range of the technique, the reconstruction doesn't work effectively. Specifically, there is an important overestimation (one order of magnitude) of the amplitude and large errors (tens of degrees) on the WNA. The ellipticity is also not well reconstructed. However, the sunward direction of propagation is clearly found.

 These three cases (Case 1, Case 2, Case 3) represent a range of whistler wave parameters that can be observed by PSP during Encounter 1. Two of them (Cases 1 and 2) are within the method applicability limits. In both cases, all important reconstructed parameters are in good agreement with the measurements. Moreover, the theoretical error based on the wave frequency and using a proxy of the propagation angle is lower than the experimental error. Case 3 is an atypical case because of its high values of $f/f_{\rm ce} \sim 0.2-0.45$, its low phase speed and, its obliquity and is out of the applicability range. In this last case, the reconstructed waveform amplitudes and the power spectral density are largely overestimated and the reconstructed WNA differs by several tens of degrees from the actual value. However, the sunward propagation direction was clearly identified. 


 \section{Application of the method to the data collected after March 2019}  

 \subsection{Discussion on the applicability of the method}
  \label{Discussion on the applicability of the method for encounter 2 and following}
For Encounter 2 and the following ones, we do not know the typical characteristics of the waves in the plasma frame. It is therefore not certain that the method is applicable in 80\% of cases as in the first encounter. On the other hand, here are some arguments that suggest that the method should work in many cases: \\
Firstly, one of the most important sources of error in the first encounters is the ratio $V_{\rm SW \perp}/ V_{\rm \varphi}$. Phase velocity increases when getting closer to the Sun and should be multiplied by about 3 at 10 $R_{\odot}$ compared to Encounter 1 \cite{bale_fields_2016}, which will greatly reduce the error. \\
Moreover, a simple Parker spiral model predicts that the background magnetic field is more radial as we get closer to the Sun. Therefore, the perpendicular component of the solar wind speed will tend to decrease. The fact that the background magnetic field is more radial should also reduce the number of configurations in which the background magnetic field is aligned with $\Vec{u}$. \\
Another important source of error is the $(\omega/\omega_{\rm ce}) \tan \theta$ term, whose evolution cannot be predicted for the next encounters. However, Encounters 2 and 3 have similar perihelion distances and the following ones will slowly approach 10 $R_{\odot}$. This suggests that for at least some perihelia the waves should have similar characteristics to those observed in the first encounter. Then, as mentioned in Section \ref{Method}, \citeA{cattell_parker_2022} statistics from 9 encounters showed that whistler waves frequency in the spacecraft frame was below 0.2 $f_{\rm ce}$ with the tendency to decrease below 0.1 approaching the Sun. Figure \ref{error_final}b shows that in the case where $\omega/ \omega_{\rm ce} \le 0.1$ this term gives an error of less than 30\% with WNAs up to 70$^\circ$.

In addition, there are several pre- and post-reconstruction methods that give indications of the quality of the reconstruction. These methods are not definitive proofs but can be used as good indicators of correct reconstruction. \\
One can use a pre-check based on $\omega/ \omega_{\rm ce}$ in the spacecraft frame (as the phase velocity increases, the measured $\omega/ \omega_{\rm ce}$ becomes closer to the one in the plasma frame). By using a proxy for the propagation angle (based on $\omega/ \omega_{\rm ce}$) and the ratio $\omega_{\rm pe}/ \omega_{\rm ce}$ it is possible to calculate the phase velocity and derive the theoretical error using Equation \ref{propagation_erreur}.  \\
Moreover, outside the range of applicability, we do not expect to reconstruct a good circular polarization (see Section \ref{Approximations}). The circular polarization can therefore be used as a post-reconstruction indicator of the method's effectiveness. 


\subsection{Application of the reconstruction technique to whistler waves recorded during Encounter 2 (no $B{\rm w}_{\rm u}$ measurements): 2019/04/03, 05:48:35-05:48:37 UT}
\label{application}
\begin{figure}[H]
\includegraphics[width=100ex]{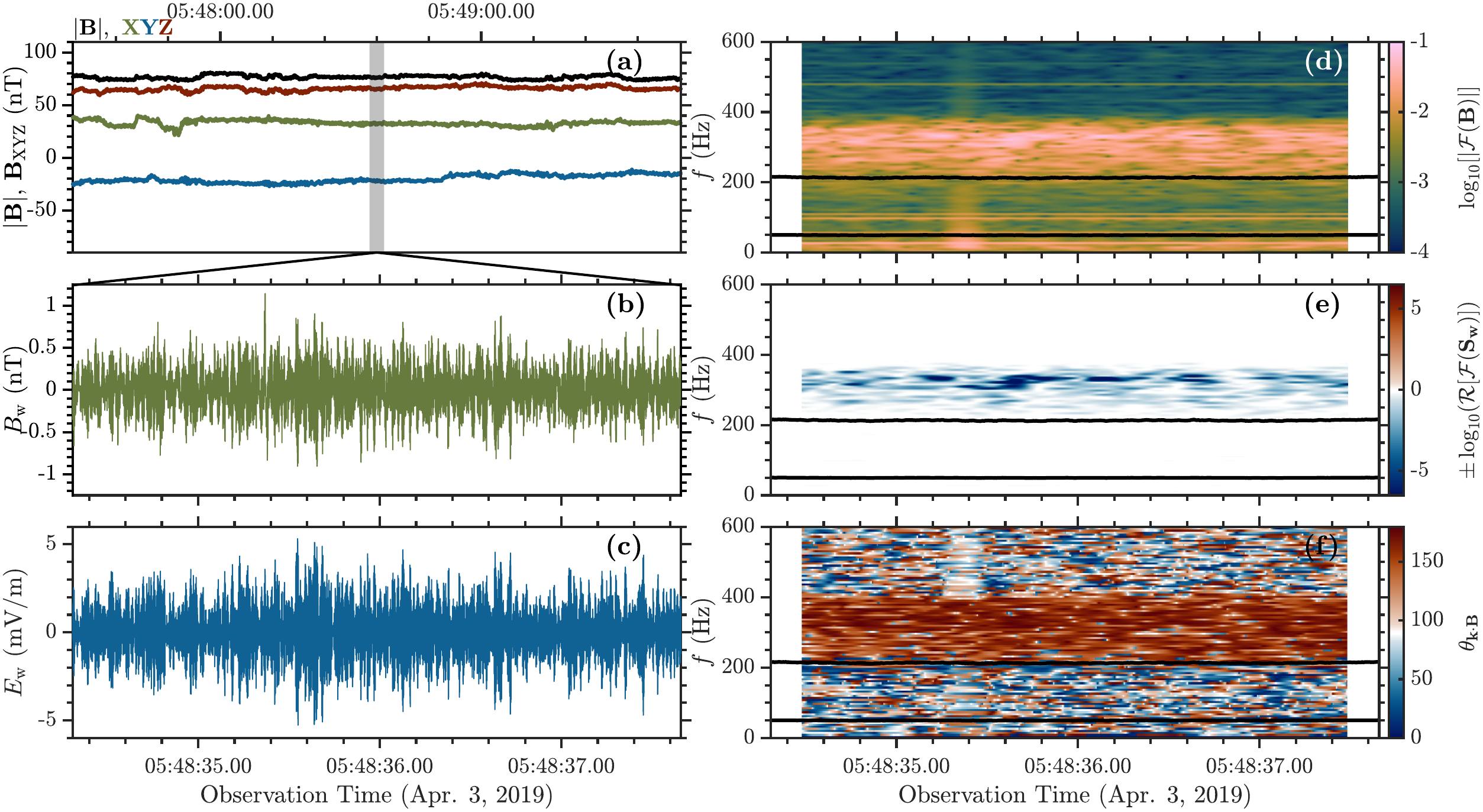}
\caption{Whistler wave packet recorded by PSP on April 3, 2019. (a) - background magnetic field in spacecraft coordinates over a short window around the burst. (b) - burst waveform of the Y magnetic field component, $B{\rm w}_{\rm Y}$. (c) - burst waveform of the Y electric field
component, $E{\rm w}_{\rm Y}$. (d) - spectrogram of the magnetic field burst waveform. (e) - Z  component of the Poynting flux. (f) - reconstructed  WNA $\theta_{\mathbf{k\cdot B}}$, ranging from $0^\circ$ to $180^\circ$ and indicating parallel (below 90$^\circ$) and anti-parallel propagation (above 90$^\circ$), respectively. The lower and upper solid lines in (d)-(f) indicate $f_{\rm lh}$ and $0.1f_{\rm ce}$, respectively. For panels (d) to (f) the frequency is shown in the spacecraft frame.}
\label{fig:application}
\end{figure}

\begin{figure}[H]
\includegraphics[width=100ex]{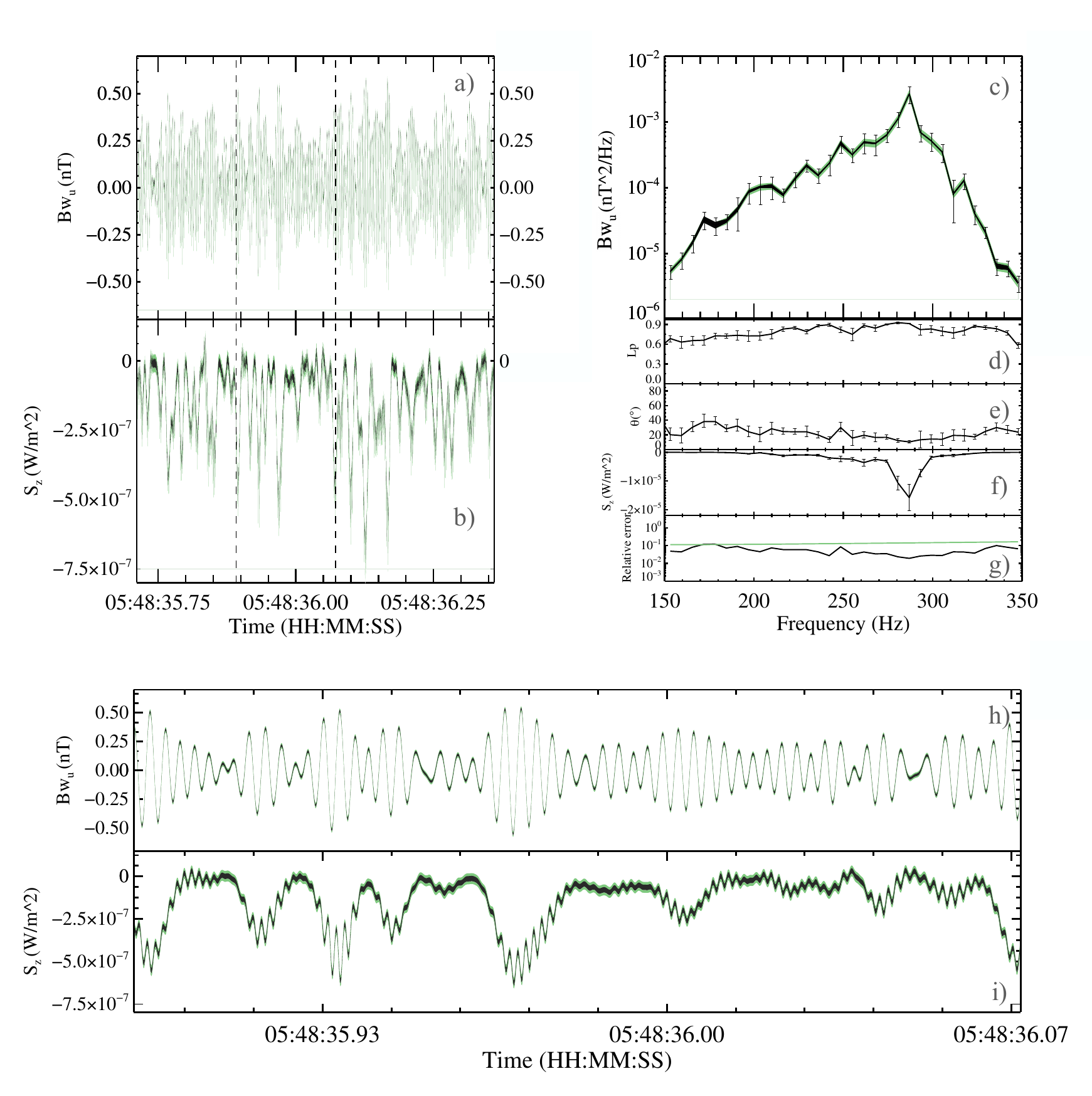}
\caption{Reconstructed whistler wave packet recorded by PSP on April 3, 2019. The error bars are detailed in Section \ref{Case1}. 
(a) - burst waveform of the u magnetic field component. (b) - Z component of the Poynting flux. The panels (c) to (g) show the frequency in the plasma frame.  
(c) - power spectral density of the u component. (d) - ellipticity. (e) - WNA $\theta_{\mathbf{k\cdot B}}$. (f) - Z component of the Poynting flux. (g) - theoretical relative error in black and using a proxy of the angle of propagation in green.  Panels (h) and (i) show a zoom on the period between the vertical dotted lines in panels (a) and (b).}
\label{fig:application2}
\end{figure}

Figures \ref{fig:application} and \ref{fig:application2} present a reconstructed whistler wave packet from Encounter 2, recorded at the heliocentric distance of $\sim37$ ($R_{\odot}$) in the FIELDS burst mode (150 $\si{\kilo \hertz}$) at 05:48:35 UTC, April 4, 2019. 
The background magnetic field magnitude is 73 nT; the plasma density is $\sim170$ $cm^3$; $(\omega_{\rm pe}/\omega_{\rm ce})^2 \sim 3250$; the bulk radial plasma velocity is 310 km/s; and $\sin(\theta_{\vec{u},\vec{B_0}}) = 0.98$. The WNA, direction of propagation, and frequency in the plasma reference frame are unknown without reconstruction due to the technical issue on the u component since March 2019. 
Figure \ref{fig:application}e and Figures \ref{fig:application2}b, \ref{fig:application2}b and \ref{fig:application2}i show that the reconstructed propagation direction is anti-sunward. The reconstructed WNA is less than 30$^\circ$ (Figure \ref{fig:application}f and Figure \ref{fig:application2}e) and the reconstructed planarity is bigger than 0.6 over the whole frequency range of the wave (Figure \ref{fig:application2}d). We can therefore deduce that $\frac{V_{\rm SW \perp}}{V^{\rm re}_{\rm \varphi}} \sim 0.06$ and $f/f_{\rm ce} \sim 0.13$ (in the plasma frame). The reconstructed components thus show that we are well within the range of application of the method and $|B^{\rm error}{\rm w}_{\rm u}  / B{\rm w}|$ is in the range of $0.03-0.1$ (Figure \ref{fig:application2}g). The method proposed in Section \ref{Discussion on the applicability of the method for encounter 2 and following} using a proxy for the propagation angle also shows that we're within the range of application (relative error less than 0.1). Moreover, ellipticity close to 1 is a good indication of correct reconstruction.


\section{Conclusions}
\label{discussion}

The Parker Solar Probe mission will last until 2025 and 24 perihelia are expected to be completed, approaching down to 10 $R_{\odot}$ and probing \textit{in situ} regions where no direct measurements have ever been made. One component of the magnetic field measured by the search-coil magnetometer is unavailable at low frequency after Encounter 1 because of a technical issue. We propose a method to determine whistler wave polarization despite the missing component. This makes it possible to evaluate wave-particle interaction effects for dynamics of the solar wind electron populations. 

To conclude:

1. We develop a technique to reconstruct the polarization parameters of whistler waves based on only two components of magnetic and electric field measurements (the PSP fields measurement regime after March 2019). We reconstruct the missing components by neglecting the electric field parallel to the background magnetic field.

2. This technique is applicable with the assumptions of (i) low-frequency
whistlers in the plasma frame relative to the electron cyclotron frequency; (ii) a small
propagation angle with respect to the background magnetic field; and (iii) a large wave
phase speed relative to the cross-field solar wind velocity. Critically, the method cannot
be applied if the background magnetic field is aligned with the affected SCM coil.

3. When within the range of applicability, all polarization parameters could be found, including propagation direction, WNA, and ellipticity. We propose pre- and post-reconstruction methods to estimate the quality of the reconstruction. One of them is to check that the ellipticity of the reconstructed magnetic field is close to 1. 

4. Our method will enable polarization properties of whistler waves in the young Solar to be determined. These polarization properties are necessary for a better understanding of particle-wave interactions.



\section{Open Research}
The data used in this work are available on the public data archive NASA CDAWeb (https://cdaweb.gsfc.nasa.gov/index.html/).

\section*{Acknowledgements}
V.K., T.D., L.C., and M.K. acknowledge funding from the CNES. L.C., V.K., and O.V.A. were supported by NASA grants 80NSSC20K0697 and 80NSSC21K1770. O.V.A and S.K. were partially supported by NSF grant number 1914670, NASA’s Living with a Star (LWS) program (contract 80NSSC20K0218), and NASA grants contracts 80NNSC19K0848, 80NSSC22K0433, 80NSSC22K0522. Parker Solar Probe was designed, built, and is now operated by the Johns Hopkins Applied Physics Laboratory as part of NASA’s Living with a Star (LWS) program (contract NNN06AA01C). We thank the FIELDS team for providing data (PI: Stuart D. Bale, UC Berkeley).

\appendix

\section{}
\label{Annexe B}
\label{Polarization}

\subsection{Estimation of the parallel electric field component in the spacecraft frame}
\label{Estimation of the parallel electric field component in the spacecraft frame}

In this section, we derive Equation \ref{Eparr_final}. 

Let us make calculations in the reference frame where $\vec{B}_{\rm 0}$ is directed along the \textbf{z} axis, and the \textbf{k}-vector is in the plane (\textbf{x}, \textbf{y}). Therefore, we have: $\mathbf{k}=k(\sin \theta \cos \varphi ,\sin \theta \sin \varphi ,\cos\theta )$, with $\varphi$ the angle between $\mathbf{x}$ and $\mathbf{k}$. To begin with, we shall treat the waves at frequencies much larger than the lower hybrid frequency (well verified for whistler waves in the solar wind) and using the cold plasma approximation. In our reference frame the dielectric tensor $\varepsilon_{\rm ij}$ reads:

\begin{equation}
\varepsilon_{\rm ij}=%
\begin{pmatrix}  
\varepsilon_{\rm 1} & i\varepsilon_{\rm 2} & 0 \\ 
-i\varepsilon_{\rm 2} & \varepsilon_{\rm 1} & 0 \\ 
0 & 0 & \varepsilon_{\rm 3}%
\end{pmatrix}
\end{equation}

where using the typical conditions of observations $\omega_{\rm pe}^{2}>>\omega _{\rm ce}^{2}>\omega ^{2}$:

$
\varepsilon _{1}=-\frac{\omega_{\rm pe}^{2}}{\omega ^{2}-\omega _{\rm ce}^{2}}%
;\varepsilon _{2}=\frac{\omega_{\rm pe}^{2}\omega _{\rm ce}}{\omega (\omega
^{2}-\omega _{\rm ce}^{2})};\varepsilon _{3}=-\frac{\omega_{\rm pe}^{2}}{\omega ^{2}}
$

One can find for whistler waves:
\[
N^{2}=\frac{\omega _{\rm pe}^{2}}{\omega (\omega_{\rm ce}\mid \cos \theta \mid
- \omega )} 
\]%
\[
\omega =\omega_{\rm ce}\mid \cos \theta \mid \frac{\frac{k^{2}c^{2}}{\omega
_{\rm pe}^{2}}}{(1+\frac{k^{2}c^{2}}{\omega _{\rm pe}^{2}})} 
\]%

The next step is to determine the polarization properties. In the following, we use $E{\rm w}(\omega,t)$ and $B{\rm w}(\omega,t)$ ($(\omega, t)$ representing the time averaged Fourier components) to approximate the theoretical fields of the general dispersion relation. We drop the $(\omega,t)$ to simplify the notations.

\begin{equation}
\begin{pmatrix}
\varepsilon_{\rm 1} - N_{\rm y}^{2} - N_{\rm z}^{2} & i\varepsilon_{\rm 2} + N_{\rm x}N_{\rm y} & N_{\rm x}N_{\rm z} \\ 
-i\varepsilon_{\rm 2} + N_{\rm x}N_{\rm y} & \varepsilon_{\rm 1}-N_{\rm x}^{2}-N_{\rm z}^{2} & N_{\rm y}N_{\rm z} \\ 
N_{\rm x}N_{\rm z} & N_{\rm y}N_{\rm z} & \varepsilon_{\rm 3} - N_{\rm x}^{2} - N_{\rm y}^{2} \\
\end{pmatrix}
\begin{array}{c}
E{\rm w}_{\rm x} \\ 
E{\rm w}_{\rm y} \\ 
E{\rm w}_{\rm z}%
\end{array}%
=0 
\end{equation}

\begin{equation}
\begin{pmatrix}
\varepsilon_{\rm 1} - N^{2}(\cos ^{2}\theta +\sin ^{2}\theta \sin ^{2}\varphi ) & 
i\varepsilon_{\rm 2} + N^{2}\sin ^{2}\theta \sin \varphi \cos \varphi & N^{2}\cos
\theta \sin \theta \cos \varphi \\ 
-i\varepsilon_{2} + N^{2}\sin ^{2}\theta \sin \varphi \cos \varphi & 
\varepsilon_{\rm 1} - N^{2}(\cos ^{2}\theta +\sin ^{2}\theta \cos ^{2}\varphi ) & 
N^{2}\cos \theta \sin \theta \sin \varphi \\ 
N^{2}\cos \theta \sin \theta \cos \varphi & N^{2}\cos \theta \sin \theta
\sin \varphi & \varepsilon_{\rm 3}-N^{2}\sin ^{2}\theta%
\end{pmatrix}
\begin{array}{c}
E{\rm w}_{\rm x} \\ 
E{\rm w}_{\rm y} \\ 
E{\rm w}_{\rm z}%
\end{array}%
=0 
\end{equation}

\begin{equation}
\begin{split}
\left(\begin{smallmatrix}
-\frac{\omega_{\rm pe}^{2}}{\omega ^{2}-\omega_{\rm ce}^{2}} - N^{2}(\cos ^{2}\theta
+\sin ^{2}\theta \sin ^{2}\varphi ) & i\frac{\omega _{\rm pe}^{2}\omega_{\rm ce}}{%
\omega (\omega ^{2}-\omega_{\rm ce}^{2})} + N^{2}\sin ^{2}\theta \sin \varphi \cos
\varphi & N^{2}\cos \theta \sin \theta \cos \varphi \\ 
-i\frac{\omega_{\rm pe}^{2}\omega_{\rm ce}}{\omega (\omega ^{2}-\omega_{\rm ce}^{2})}%
+N^{2}\sin ^{2}\theta \sin \varphi \cos \varphi & -\frac{\omega _{\rm pe}^{2}}{%
\omega ^{2}-\omega_{\rm ce}^{2}}-N^{2}(\cos ^{2}\theta +\sin ^{2}\theta \cos
^{2}\varphi ) & N^{2}\cos \theta \sin \theta \sin \varphi \\ 
N^{2}\cos \theta \sin \theta \cos \varphi & N^{2}\cos \theta \sin \theta
\sin \varphi & -\frac{\omega_{\rm pe}^{2}}{\omega ^{2}}-N^{2}\sin ^{2}\theta%
 \end{smallmatrix}\right)
\begin{array}{c}
E{\rm w}_{\rm x} \\ 
E{\rm w}_{\rm y} \\ 
E{\rm w}_{\rm z}%
\end{array}
=0 
\end{split}
\end{equation}

Polarization vectors can be expressed in the reference frame determined at the beginning as: 

\begin{equation}
\vec{E}{\rm w}=a
\begin{pmatrix}
 \frac{\omega }{(\omega ^{2}-\omega_{\rm ce}^{2})}(i\omega_{\rm ce}\sin
\varphi +\omega \cos \varphi )+\frac{k^{2}c^{2}}{\omega_{\rm pe}^{2}}\cos
\varphi  \\ 
 \frac{\omega }{\omega ^{2}-\omega _{\rm ce}^{2}}(\omega \sin \varphi
-i\omega _{\rm ce}\cos \varphi )+\frac{k^{2}c^{2}}{\omega_{\rm pe}^{2}}\sin \varphi 
\\ 
\frac{\frac{k^{2}c^{2}}{\omega_{\rm pe}^{2}}\cos \theta \sin \theta }{(1+\frac{%
k^{2}c^{2}}{\omega_{\rm pe}^{2}}\sin ^{2}\theta )}(\frac{\omega ^{2}}{(\omega
^{2}-\omega _{\rm ce}^{2})}+\frac{k^{2}c^{2}}{\omega_{\rm pe}^{2}})%
\end{pmatrix}
\end{equation}

\begin{equation}
\vec{B}{\rm w}=a\frac{ck}{\omega }%
\begin{pmatrix}
-\frac{\cos \theta \sin \varphi }{(1+\frac{k^{2}c^{2}}{\omega_{\rm pe}^{2}}%
\sin ^{2}\theta )}(\frac{\omega ^{2}}{(\omega ^{2}-\omega _{\rm ce}^{2})}+\frac{%
k^{2}c^{2}}{\omega_{\rm pe}^{2}})+i\frac{\omega _{\rm ce}\omega \cos \theta \cos
\varphi }{\omega ^{2}-\omega _{\rm ce}^{2}} \\ 
\frac{i\omega _{\rm ce}\omega \cos \theta \sin \varphi }{(\omega ^{2}-\omega
_{e}^{2})}+(\frac{\omega ^{2}}{(\omega ^{2}-\omega _{\rm ce}^{2})}+\frac{%
k^{2}c^{2}}{\omega_{\rm pe}^{2}})\frac{\cos \theta \cos \varphi }{(1+\frac{%
k^{2}c^{2}}{\omega_{\rm pe}^{2}}\sin ^{2}\theta )} \\ 
-i\frac{\omega _{\rm ce}\omega \sin \theta }{\omega ^{2}-\omega _{\rm ce}^{2}}%
\end{pmatrix}
\end{equation}

Where $a$ is a constant. 
Using the refractive index magnitude, one can re-write wave polarization dependence upon characteristic frequencies as follows:

\begin{equation}
\vec{E}{\rm w}=a%
\begin{pmatrix}
\frac{\omega }{(\omega ^{2}-\omega _{\rm ce}^{2})}(i\omega _{\rm ce}\sin
\varphi +\omega \cos \varphi )+\frac{\omega }{\omega _{\rm ce}\cos \theta -\omega 
}\cos \varphi \\ 
\frac{\omega }{\omega ^{2}-\omega _{\rm ce}^{2}}(\omega \sin \varphi
-i\omega _{\rm ce}\cos \varphi )+\frac{\omega }{\omega _{\rm ce}\cos \theta -\omega }%
\sin \varphi  \\ 
\frac{\omega \sin \theta }{(\omega _{\rm ce}-\omega \cos \theta )}(\frac{\omega
^{2}}{(\omega ^{2}-\omega _{\rm ce}^{2})}+\frac{\omega }{\omega _{\rm ce}\cos \theta
-\omega })%
\end{pmatrix}
\end{equation}

\begin{equation}
\vec{B}{\rm w}=a\frac{ck}{\omega }%
\begin{pmatrix}
-\frac{(\omega _{\rm ce}\cos \theta -\omega )\sin \varphi }{(\omega _{\rm ce}-\omega
\cos \theta )}(\frac{\omega ^{2}}{(\omega ^{2}-\omega _{\rm ce}^{2})}+\frac{%
\omega }{\omega _{\rm ce}\cos \theta -\omega })+i\frac{\omega _{\rm ce}\omega \cos
\theta \cos \varphi }{\omega ^{2}-\omega _{\rm ce}^{2}} \\ 
\frac{i\omega _{\rm ce}\omega \cos \theta \sin \varphi }{(\omega ^{2}-\omega
_{e}^{2})}+(\frac{\omega ^{2}}{(\omega ^{2}-\omega _{\rm ce}^{2})}+\frac{\omega }{%
\omega _{\rm ce}\cos \theta -\omega })\frac{(\omega _{\rm ce}\cos \theta -\omega )\cos
\varphi }{(\omega _{\rm ce}-\omega \cos \theta )} \\ 
-i\frac{\omega _{\rm ce}\omega \sin \theta }{\omega ^{2}-\omega _{\rm ce}^{2}}%
\end{pmatrix}
\end{equation}

Now one can come to simplifications. We have already noted that the
overwhelming majority of observed waves satisfy the condition $\omega<<\omega _{\rm ce},$ that is, the parameter $\frac{\omega }{\omega _{\rm ce}}=\epsilon$ is small. This allows one to use it as the small parameter constructing solutions as power series over this parameter. The first order approximation on $\epsilon $ results in:

\begin{equation}
\vec{E}{\rm w}=a%
\begin{pmatrix}
\frac{\omega }{\omega_{\rm ce}}(-i\sin \varphi +\frac{\cos \varphi }{\cos \theta 
}) \\ 
\frac{\omega }{\omega_{\rm ce}}(i\cos \varphi +\frac{\sin \varphi }{\cos \theta }%
) \\ 
\frac{\omega^2 \sin \theta}{\omega_{\rm ce}^2 \cos \theta} 
\label{E_plasma_frame}
\end{pmatrix}
\end{equation}

\begin{equation}
\vec{B}{\rm w}=a \frac{ck}{\omega }
\begin{pmatrix}
-\frac{\omega}{\omega_{\rm ce}}(\sin \varphi +i\cos \theta \cos \varphi ) \\ 
\frac{\omega}{\omega_{\rm ce}}(-i\cos \theta \sin \varphi +\cos \varphi )  \\ 
i\frac{\omega}{\omega_{\rm ce}}\sin \theta 
\end{pmatrix}
\end{equation}

The electric field is measured in the spacecraft frame, which is different from the plasma frame. It is therefore necessary to take Lorentz transformations into account. For the magnetic field, since the measured solar wind speed ($V_{\rm SW}$) verifies $V_{\rm SW} << c$ (where $c$ is the speed of light) these transformations can be neglected and we can safely consider that $\vec{B}^{\rm SC}{\rm w} = \vec{B}{\rm w}$. 
For the electric field, on the other hand, we have: 

\begin{equation}
\vec{E}^{\rm SC}{\rm w} = \vec{E}{\rm w} - \frac{1}{c} \bigl( \Vec{V}_{\rm SW} \times \vec{B}{\rm w}  \bigr)
\end{equation}

where $\Vec{V}_{\rm SW}$ is expressed as: 

\begin{align}
\Vec{V}_{\rm SW}= \begin{array}{c}
V_{\rm SW x}\\ 
V_{\rm SW y} \\ 
V_{\rm SW z}%
\end{array} 
\end{align}

\begin{equation}
\frac{1}{c} \bigl( \Vec{V}_{\rm SW} \times \vec{B}{\rm w}  \bigr)= a\frac{k}{\omega} %
\begin{pmatrix}
\frac{\omega}{\omega_{\rm ce}}(i \sin \theta V_{\rm SW y} - V_{\rm SW z}(-i \cos \theta \sin \varphi + \cos \varphi)) \\ 
\frac{\omega}{\omega_{\rm ce}} (- V_{\rm SW z}( \sin \varphi + i \cos \theta \cos \varphi) - i \sin \theta V_{\rm SW x}) \\ 
\frac{\omega}{\omega_{\rm ce}}(V_{\rm SW x}( -i \cos \theta \sin \varphi + cos \varphi) + V_{\rm SW y} ( \sin \varphi + i \cos \theta \cos \varphi ))
\end{pmatrix}
\end{equation}

Therefore, 
\begin{equation}
|\frac{1}{c} \bigl( \Vec{V}_{\rm SW} \times \vec{B}{\rm w}  \bigr)_{\rm z}| =  a \frac{k}{\omega} \frac{\omega}{\omega_{\rm ce}} \bigl( (V_{\rm SW x} \cos \varphi + V_{\rm SW y} \sin \varphi )^2 + \cos^2 \theta(  V_{\rm SW y} \cos \varphi - V_{\rm SW x} \sin \varphi)^2 \bigr)^{1/2} 
\end{equation}

By choosing, $\varphi = 0$, then

\begin{equation}
|\frac{1}{c} \bigl( \Vec{V}_{\rm SW} \times \vec{B}{\rm w}  \bigr)_{\rm z}| =  a \frac{k}{\omega} \frac{\omega}{\omega_{\rm ce}} \bigl( V_{\rm SW x}^2 + V_{\rm SW y}^2 \cos^2 \theta \bigr)^{1/2} 
\end{equation}

Thus, 

  \begin{equation}
        |E^{\rm SC}\rm w_{||}| 	\leq |E{\rm w}|  (\frac{V_{\rm SW\perp}}{V_{\varphi}} + (\frac{\omega}{\omega_{\rm ce}}) \tan{\theta} )
   \label{Eparr_final_appendix}
 \end{equation}

\subsection{Propagation of the error}
\label{Propagation of the error}

The electromagnetic wave equation is expressed as:

  \begin{equation}
          \Vec{E}{\rm w} \cdot \Vec{B}{\rm w} = 0
\end{equation}

and is also valid in the spacecraft frame: 

  \begin{equation}
          \Vec{E}^{\rm SC}{\rm w} \cdot \Vec{B}^{\rm SC}{\rm w} = 0
\end{equation}

Therefore, 

  \begin{equation}
          \Vec{E}^{\rm SC}{\rm w ||} \cdot \Vec{B}^{\rm SC}{\rm w ||} + \Vec{E}^{\rm SC}{\rm w \perp} \cdot \Vec{B}^{\rm SC}{\rm w \perp}= 0
\end{equation}

In our approximation we neglect $\Vec{E}^{\rm SC}{\rm w ||}$, therefore using results from the previous section we find

  \begin{equation}
          |\Vec{E}^{\rm SC re}{\rm w} \cdot \Vec{B}^{\rm SC}{\rm w}| \leq |E{\rm w}|  (\frac{V_{\rm SW\perp}}{V_{\varphi}} + (\frac{\omega}{\omega_{\rm ce}}) \tan{\theta} ) \sin{\theta}
\end{equation}



One way of approximating the error on the reconstructed component, considering these two errors as independent and following a normal distribution, is as follows:

\begin{equation}
           ( | B^{\rm error}{\rm w}_{\rm u}  / B{\rm w}| ) \lesssim \sqrt{( \frac{V_{\rm SW\perp}}{V_{\varphi}})^2 + (  (\frac{\omega}{\omega_{\rm ce}}) \tan{\theta})^2 }  \frac{\sin{\theta} }{ \sin{\theta_{\Vec{B_{\rm 0}}, \Vec{u}}}} 
\end{equation}

\section{}
\label{SC_SCM}
\begin{figure}[H]
\includegraphics[width=100ex]{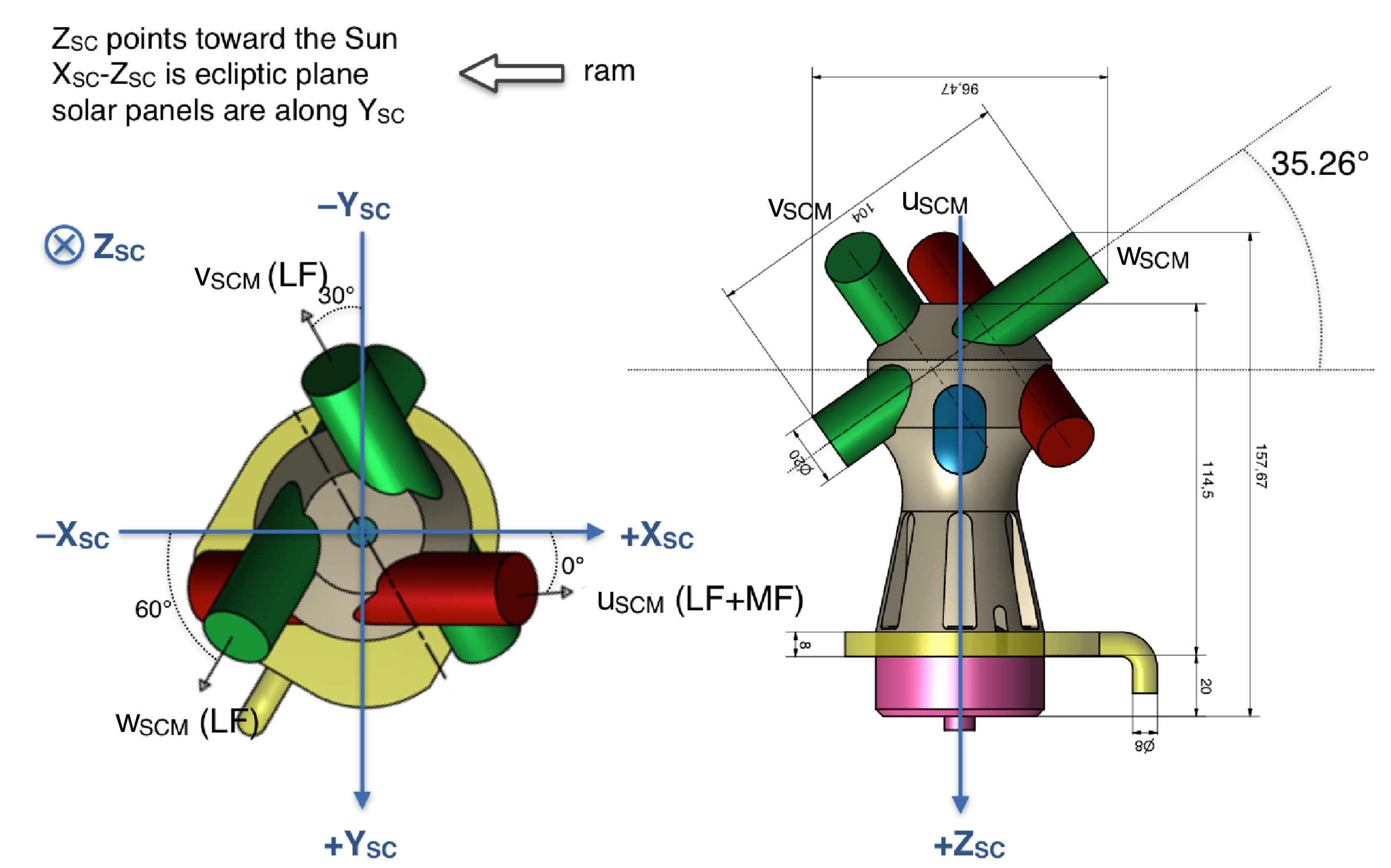}
\caption{Schematics of the SCM and of the relation between its reference frame (\textbf{u},\textbf{v},\textbf{w}) and the one of the spacecraft (\textbf{X},\textbf{Y},\textbf{Z}). 
}
\label{SCM_frame}
\end{figure}

Figure \ref{SCM_frame} represents the relationship between the SCM and the spacecraft reference frame. The rotation matrix between these two frames is the following: 

\begin{equation}
R_{{\rm SCM \rightarrow SC}_{\rm ij}}=
\begin{pmatrix}
0.81654 & -0.40827 & -0.40827 \\ 
0 & -0.70715 & 0.70715\\ 
-0.57729 & -0.57729 & -0.57729 \\
\end{pmatrix}
\end{equation}

\color{black}

\bibliography{papier_reconstruction2}

%
%
%
%
%

\end{document}